\documentclass{aa}

\usepackage{txfonts}
\usepackage{graphicx}
\usepackage{natbib}
\bibpunct{(}{)}{;}{a}{}{,} % to follow the A&A style\usepackage{mathcomp}

\begin{document}

\title{The HARPS search for southern extra-solar planets. XXXV. 
  Super-Earths around the M-dwarf neighbors Gl~433 and Gl~667C 
\thanks{Based on observations collected with HARPS instrument on the
  3.6-m telescope at La Silla Observatory (European Southern
  Observatory) under programs ID072.C-0488(E)}} 
 
\subtitle{}

\author{X. Delfosse \inst{1}
\and X. Bonfils \inst{1}
\and T. Forveille \inst{1}
\and S. Udry \inst{2}
\and M. Mayor \inst{2}
\and F. Bouchy \inst{3}
\and M. Gillon \inst{4}
\and C. Lovis \inst{2}
\and V. Neves \inst{1,5,6}
\and F. Pepe \inst{2}
\and C. Perrier \inst{1}
\and D. Queloz \inst{2}
\and N.C. Santos \inst{5,6}
\and D. S\'egransan \inst{2}
}
\institute{UJF-Grenoble 1 / CNRS-INSU, Institut de Plan\'etologie et
  d'Astrophysique de Grenoble (IPAG) UMR 5274, Grenoble, F-38041,
  France  
\and
           Observatoire de Gen\`eve, Universit\'e de Gen\`eve, 
           51 ch. des Maillettes, 1290 Sauverny, Switzerland
\and
           Institut d'Astrophysique de Paris, CNRS, Universit\'e Pierre
           et Marie Curie, 98bis Bd. Arago, 75014 Paris, France 
\and 
           Université de Li\`ege, All\'ee du 6 ao\^ut 17, Sart Tilman,
           Li\`ege 1, Belgium  
\and
           Centro de Astrof\'isica, Universidade do Porto,
           Rua das Estrelas, 4150-762 Porto, Portugal 
\and
           Departamento de F\'{\i}sica e Astronomia, Faculdade de Ci\^encias,
           Universidade do Porto, Rua do Campo Alegre, 4169-007 Porto,
           Portugal
 }

\date{Received / Accepted}
\abstract{M dwarfs have been found to often have super-Earth planets
with short orbital periods. Such stars are thus preferential
targets in searches for rocky or ocean planets in the solar
neighbourhood.}
{In a recent paper (Bonfils et al. 2011), we
announced the discovery of respectively 1 and 2 low mass planets 
around the M1.5V stars Gl~433 and Gl~667C. We found those planets
with the HARPS spectrograph on the ESO~3.6-m telescope at La Silla 
Observatory, from observations obtained during the Guaranteed Time 
Observing program of that instrument.}
{We have obtained additional HARPS observations of those two stars,
for a total of respectively 67 and 179 Radial Velocity measurements
for Gl~433 
and Gl~667C, and present here an orbital analysis of those 
extended data sets and our main conclusion about both planetary
systems.}
{One of the three planets, Gl~667Cc, has a mass of only
  $M_2.\sin{i} \sim  4.25~M_{\oplus}$ and orbits in the central
habitable zone of its host star. It receives just 10\% less stellar
energy from Gl~667C than the Earth receives from the Sun. However
planet evolution in habitable zone can be very different if the host
star is a M dwarf or a solar-like star, without necessarily questioning
the presence of water. The two other
planets, Gl~433b and Gl~667Cb, both have $M_2.\sin{i}$ of 
$\sim$5.5~$M_{\oplus}$ and periods of $\sim$7 days. The
Radial Velocity measurements of both stars contain longer time scale
signals, which we fit as longer period Keplerians. For
Gl~433 that signal probably originates in a Magnetic Cycle, 
while a longer time span will be needed to conclude for Gl~667C. 
The metallicity
of Gl~433 is close to solar, while Gl~667C is metal poor with
[Fe/H]$\sim$-0.6. This reinforces the recent conclusion that
the occurence of Super-Earth planets does not strongly correlate with
stellar metallicity.}{}  


%\titlerunning{}

\authorrunning{Delfosse et al.}
\titlerunning{Super-Earths around Gl~433 and Gl~667C}
\maketitle

\section{Introduction}

Much interest has recently focused on planets around M dwarfs,
with three main motivations: constraining formation, 
physico-chemical characterization of planets, and finding
rocky planets in the habitable zone of their stars. The compared
occurence frequency, as a function of orbital elements, of 
planets around M dwarfs and around the more massive solar-type 
stars probes the sensitivity of planetary formation to its 
initial conditions. Giant planets are rare around M~dwarfs, 
with \citet{bonfils2011}, for instance, finding a low frequency 
of $6^{+6}_{-2}$\% for periods under 10000~days. That number is
lower than the  
$~10\pm2$\% frequency for similar planets around solar-like stars 
\citep[e.g.][]{mayor2011}, though not yet at a high significance level. 
Super-Earth (2-10M$_{\oplus}$), by contrast, seem
abundant around M dwarfs at the short orbital periods to which
radial velocity searches are most sensitive: \citet{bonfils2011} 
find an occurence rate of $88^{+55}_{-19}$\% for P$<$10 days, to be 
compared to $\sim50$\% for similar planets around G dwarfs.

The physical conditions in the circumstellar disks of very low mass
stars therefore favor the formation of low-mass planets (rocky and
maybe ocean planets) close to the star. This translates into
good odds for finding telluric planets transiting M dwarfs
\citep[like those very recently announced
  by][]{muirhead2012}.  
Such transiting planets are excellent atmospheric characterization 
targets: since transit depth scales with the squared stellar radius,  
a transit across an M dwarf provides much more accurate measurements 
of radius and transmission spectrum than across a solar-type star.

Finding rocky planets in the habitable zone (HZ) of their stars is 
another motivation for planet searches around M dwarfs. Planets of 
given mass and orbital separation induce larger stellar radial velocity 
variation around lower mass stars, but, more importantly, the low 
luminosity of M dwarfs moves their habitable zone much nearer to 
the star. These two effects combine, and a habitable planet around 
a 0.3-M$_{\odot}$ M~dwarf produces a 7 times larger radial velocity wobble
than the same planet orbiting a solar-mass G dwarf. Additionally,
some atmospheric models suggest additional advantages of planets
in the HZ of M dwarfs for habitability characterization: 
\citet{segura2005} find that the significant dependence of
atmospheric photochemistry on the incoming spectral energy 
distribution strongly reinforces some biomarkers 
in the spectra of Earth-twin planets orbiting M dwarfs.
N$_2$O and CH$_3$Cl, in particular, would be detectable 
for Earth-twin planets around M dwarfs but not around 
solar-type stars.

As discussed in \citet{bonfils2011}, which presents the full 
specifications of the survey, we have been monitoring the radial 
velocities (RV) of a distance and magnitude limited sample of 102 
M dwarfs since 2003 with the HARPS spectrograph mounted on 
the ESO/3.6-m La Silla telescope. With a typical RV accuracy of 1-3m/s
and 460 hours of observations, our survey identified super-Earth and 
Neptune-like planets around Gl~176 \citep{forveille2009}, Gl~581 
\citep{bonfils2005a, udry07, mayor09} and Gl~674 \citep{bonfils2007}. 
In \citet{bonfils2011}, which was centered on the statistical 
implications of the survey for planet populations, we added to 
this list 1 planet around Gl~433 and 2 around Gl~667C, providing 
detailed periodograms and a false alarm probability in using our
GTO/HARPS data. \citet{anglada2012} recently announced 
a confirmation of the planets Gl~667Cb and Gl~667Cc, which however
is only very partly independent, since it largely rests on our
\citet{bonfils2011} observations and data reduction.

Here we present a more detailed orbital analysis of these two
systems than we could present in \citet{bonfils2011}, and we refine
their parameters by adding new seasons of RV measurements. 
The 3 super-Earths have $M_2.\sin{i}$ between 4.25 and 5.8
M$_{\oplus}$. With periods of $\sim$7days, both Gl433b and Gl667Cb 
are hot super-Earths. Gl667Cc, by contrast, has a 28-day period.
It orbits in the centre of the habitable zone of its star, and 
receives just 10\% less stellar energy than the Earth receives 
from the Sun. We also detect some longer-period radial-velocity
variations, which we also discuss.

The next section discusses our data taking and analysis, while
Sect.~3 summarizes the stellar characteristics of Gl~433 and
Gl~667C. Sect.~4 and 5 respectively present our orbital analyses 
of the planetary systems of Gl~433 and Gl~667C. Sect.~6 discusses
the habitable zones of M dwarfs, with emphasis on the case of 
Gl~667Cc. Finally, Sect.7. summarizes our conclusions.

\section{Spectroscopic and Doppler measurement with HARPS}

Our observing procedure is presented in some details in
\citet{bonfils2011}, and is only summarized here. For both stars we
obtained 15 min exposures with the HARPS spectrograph \citep[High
  Accuracy Radial velocity Planets Searcher][]{mayor2003}. HARPS is
a fixed-format echelle spectrograph, which covers the 380 to 630~nm 
spectral range with a resolving power of 115 000. HARPS is
fed by a pair of fibres, and is optimized for high accuracy
radial-velocity measurements, with a stability better than 1m/s
during one night. To avoid light pollution on the stellar spectrum of our
``faint'' M-dwarf targets, we chose to keep dark the calibration
fiber of the spectrograph. Our Radial Velocity accuracy is 
therefore intrinsically limited by the instrumental stabiliy of
HARPS. That stability is however excellent and the signal-to-noise ratio 
of our 15min exposures of these two V$\sim$10 M~dwarfs 
limits the RV precision to slightly above 1~m/s (the median S/N
ratio per pixel at 550nm is 57 and 65 respectively for Gl~433 and
Gl~667C). Sect.~4 and 5 discuss the number of exposures and 
their time span for the two stars.

For homogeneity, we reprocessed all spectra with the latest
version of the standard HARPS pipeline. The pipeline \citep{lovis2007}
uses a nightly set of calibration exposures to locate the orders,
flat-field the spectra (Tungsten lamp illumination) and 
to precisely determine the wavelength calibration scale
(ThAr lamp exposure). We measured the Radial Velocity through
cross correlation of the stellar spectra with a numerical weighted
mask, following the procedure of \citet{pepe2002}. For both
Gl~433 and Gl667C, we used a mask derived from a very high S/N spectrum 
of a M2 dwarf. The velocities of Gl~433 and Gl~667C have internal median
errors of respectively 1.15 and 1.30~m/s. This includes the
uncertainty of the nightly zero point calibration, the drift
and jitter of the wavelength scale during a night, and the
photon noise (the dominant term here).

Because both stars have significant proper motion, the projection 
of their velocity vector changes over the duration of our survey. 
We subtract this secular acceleration \citep[see][for details]{kurster2003} 
before radial velocity analysis. For Gl~433 and Gl~667C the values are
respectively 0.15 and 0.21~m/s/yr.

\section{Stellar characteristic of Gl~433 and Gl~667C }

Both Gl~433 (LHS~2429) and Gl~667C are early-M dwarfs in the 
close solar neighborhood. Table~\ref{tab_parametre} summarizes 
their properties. The masses are computed from the $K$-band absolute 
magnitudes using the \citet{delfosse00} empirical near-infrared 
mass-luminosity relation. The bolometric correction of Gl~433, and 
then its luminosity, is computed from the $I-K$ color using 
the cubic polynomial of \citet{leggett2000}. Those authors 
directly determined the luminosity of Gl~667C using a combination
of flux calibrated observed spectra and synthetic spectra, and we
adopt their value.  

\begin{figure}
\includegraphics[width=9cm,angle=0]{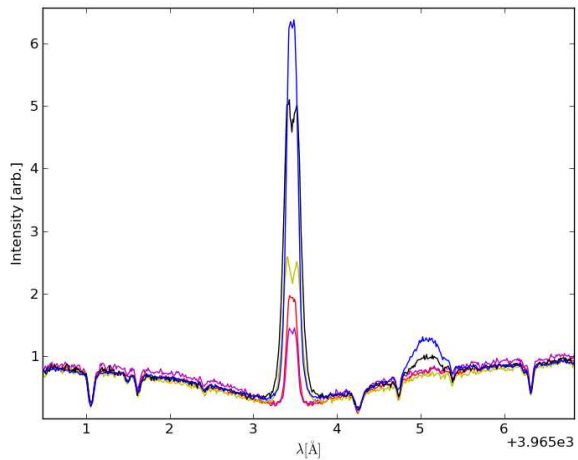}
\caption{
Emission reversal in the Ca II H line in the average spectra of (from
the less to the more active) Gl~581 (M3, in magenta), Gl~667C (M1.5,
in pink), Gl~433 (M1.5, in yellow), Gl~176(M2.5, dark) and Gl~674 (M3,
in blue). Within our 100 M dwarfs
sample, Gl 581 has one of the weakest Ca II emission and illustrates a
very quiet M dwarf. Gl~674 and Gl~176 have a much stronger emission and
are both moderately active with an identified rotational period
of respectively 35 and 39 days \citep{bonfils2007, forveille2009}.
}
\label{calcium}
\end{figure}

\begin{table}
\begin{tabular}{lll} \hline
 & Gl~433 & Gl~667C \\ \hline \hline
Spectral type$^{(1)}$   & M1.5 & M1.5 \\
V              &  9.79 & 10.22 \\
J$^{(2)}$       &  6.47$\pm$0.02 & 6.85$\pm$0.02 \\
H$^{(2)}$       &  5.86$\pm$0.04 & 6.32$\pm$0.04 \\
K$^{(2)}$       &  5.62$\pm$0.02 & 6.03$\pm$0.02 \\
$\pi$          & 110.6$\pm$1.8$^{(3)}$ & 138.2$\pm$0.7$^{(4)}$ \\ 
M (in M$_\odot$)$^{(5)}$ & 0.48 & 0.33 \\
L (in L$_\odot$)$^{(6)}$ & 0.034 & 0.014 \\
$[$Fe/H$]$$^{(7)}$  & -0.22 & -0.55 \\
T$\rm{eff}$$^{(8)}$ &  3600 & 3600 \\ 
$\log{R_X}$$^{(9)}$ & $<$-4.8 & $<$-4.12 \\ \hline
\end{tabular}
\caption{(1) \citet{hawley96}; (2) \citet{cutri2003}; (3)
  \citet{hipparcos1997}; (4) \citet{soder1999}; (5) from
  mass-luminosity relation of \citet{delfosse00} ;
  (6) \citet{leggett2000}; (7) from luminosity-color relation of
  \citet{bonfils2005}; (8) \citet{morales08}; (9) from
  \citet{schmitt1995} for Gl~433 and from \citet{schmitt2004} for Gl~667C}
\label{tab_parametre}
\end{table}

\subsection{Gl~433 : metallicity, activity and dynamic population}

According to \citet{bonfils2005} we estimated the metallicity of Gl~433
to [Fe/H]$\sim$-0.2. This photometric calibration relationship is
based on a faint number of stars in the high metallicity range 
but \citet{neves2012} show that for such value of [Fe/H] the
calibration is correct with a typical dispersion of 0.2~dex. This is
confirmed by the value of [Fe/H]$=$-0.13 obtained in using
\citet{neves2012} relationship, itself an update from
\citet{schlaufman2010}. We could
conclude that Gl~433 is solar or slightly sub-metallic.

Gl~433 is dynamically classified as a membership of old disk
\citep{leggett1992}. Ca~II H and K chromospheric emission is
determined by \citet{rauscher2006} and Gl~433 is in the less active
half of M dwarfs with same luminosity. For objects of
similar spectral type a direct comparison of Ca~II emission lines gives
relative estimation of rotational period (larger Ca~II emissions
correspond to shorter rotational periods). In the Fig~\ref{calcium} we
present an average HARPS spectrum of Gl~433 in the region of Ca~II H
line. Gl~433 has an emission of Ca~II slightly higher than Gl~581 (a very
quite M dwarf), but well below those of Gl~176 and Gl~674 for whose
rotational periods of respectively 35 and 39 days have been determined
by \citet{bonfils2007} and \citet{forveille2009}. The five M dwarfs of the
Fig.~\ref{calcium} do not have identical spectral (from
M1.5 to M3) therefore estimation of rotation from the Ca~II emission
is only indicative. It shows that Gl~433 rotates most probably with a
period longer than Gl~176 and Gl~674.

The X-ray flux of Gl~433 is not 
detected by ROSAT and we use a ROSAT limiting sensitivity of
$2.5\time10^{27}.[d/10pc]^2$~erg/s \citep{schmitt1995} to estimate
$R_X=Log{L_X/L_{BOL}}<-4.8$. For an M dwarfs of $\sim0.5$M$_{\odot}$
the $R_X$ versus rotation period relation of \citet{kiraga2007} give
$P_{rot}~>~40$~days for such level of X flux. This inferior limit on
the rotation period is coherent with the estimate from the Ca~II
emission.

\subsection{Gl~667C : multiplicity, metallicity, activity and dynamic
  population} 

Gl~667C is the lightest and isolated component of a hierarchical
triple system, the two others components are a closest
couple of K dwarfs. Gl~667AB has a semi-major axis of
1.82~A.U. (period of 42.15 years)
and a total mass dynamically determined of 1.27M$_{\odot}$
\citep{soder1999}. Gl~667C is at a projected distance of 32.4'' of
Gl~667AB, giving an expected semi-major axis of $\sim$300~A.U. \citep[for
a distance of 7.23~pc and a factor of 1.26 between expected and
projected semi-major axis,][]{fischer1992}.

The \citet{bonfils2005} photometric relationship for Gl667C gives us
an estimate of the metallicity of ~0.55 dex, which agrees quite well
with the spectroscopic determination of
[Fe/H]$\sim$-0.6 for the primary \citep{perrin1988,zakho1996}. This
demonstrates that the \citet{bonfils2005} photometric calibration gives
excellent results for low-metallicity M dwarfs. The
\citet{neves2012} relationship gives [Fe/H]$\sim$-0.45, confirming
that this star is a metal-poor M dwarfs.

Consistent with this chemical composition, Gl667 is classified as
member of the old
disk population from its UVW velocity \citep{leggett1992} and is among
the objects of lower chromospheric emission for its luminosity
\citep{rauscher2006}. In the Fig.~\ref{calcium} the Ca~II H line
emission of Gl~667C is slightly inferior to the Gl~433 one, also indicating
also a large rotational period. 

The NEXXUS database
\citep{schmitt2004} indicates detection of the X-ray emission of
Gl~667C from the ROSAT All-Sky Survey (RASS) with a value of
$\log{L_X}=27.89$. But the resolution of ROSAT image is not sufficient
to separate the three components of the system. Although this value is the
jointed flux for Gl~667ABC, it gives a superior limit on the coronal
emission of Gl~667C. This implies $R_X=Log{L_X/L_{BOL}}$ well below
$-4.12$ and rotational period superior to 40~days (in applying the
\citet{kiraga2007} $R_X$ versus rotation period relationship).

\section{Orbital analysis of Gl~433}

\subsection{HARPS measurements}

We obtained 67 measurements of Gl433 radial velocity that span 2904
days from December 2003 to November 2011. The data add 17 new points to
those analyzed in \citet{bonfils2011} and extend their time span by
1200 days. We 
estimate that the uncertainties of $\sim$1.1 m/s are dominated by photon
noise (the median S/N ratio per pixel at 550 nm is 57). Overall, the
RVs have a rms=3.14 m/s and $\sqrt{\chi^2}=7.7\pm0.5$ per degree of
freedom, confirming variability in excess to measurement
uncertainties ($\sim$1.1 m/s, dominated by photon noise).  

\begin{figure}
\begin{tabular}{c}
\includegraphics[width=9cm,angle=0]{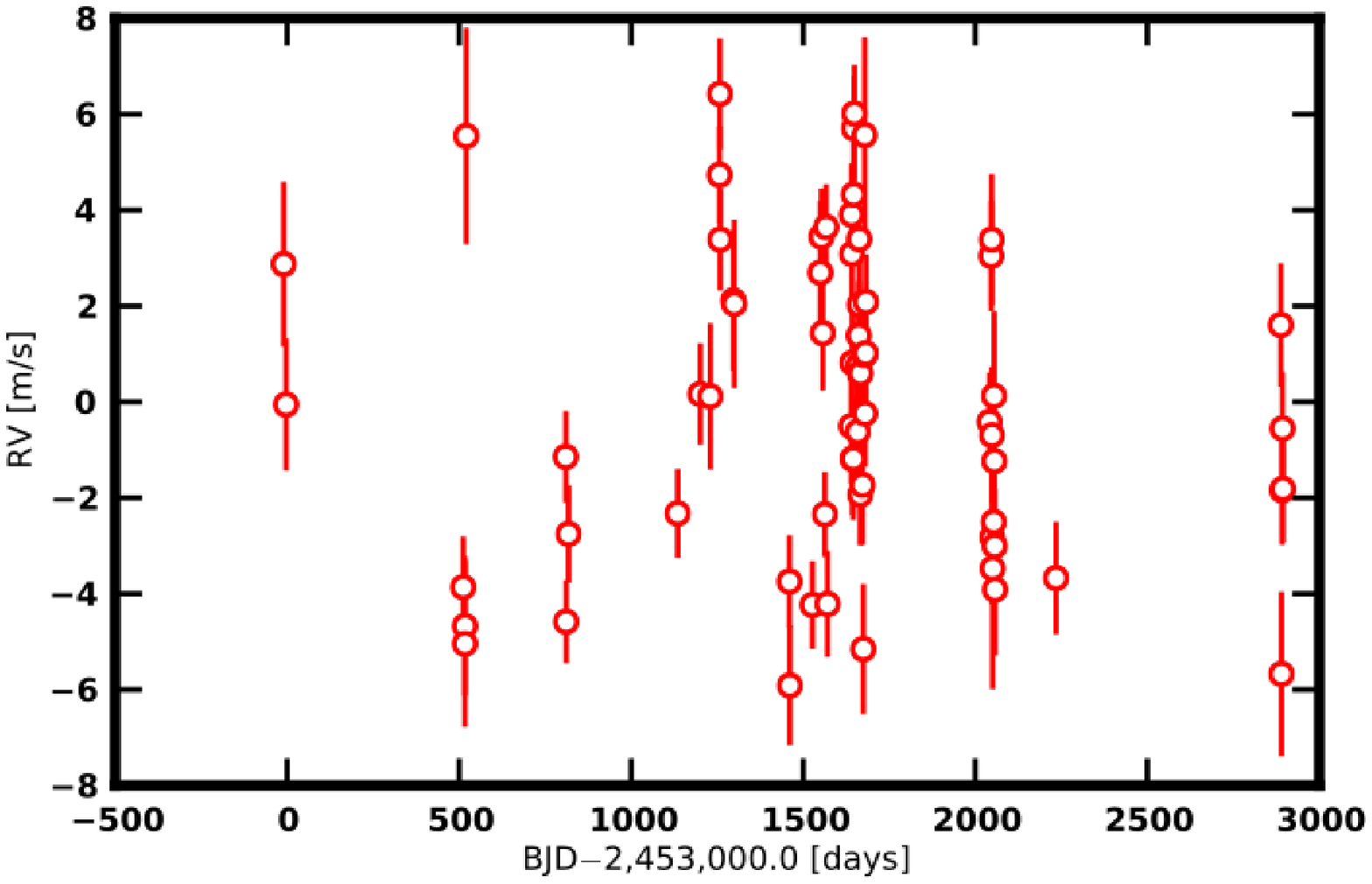} \\
\includegraphics[width=9cm,angle=0]{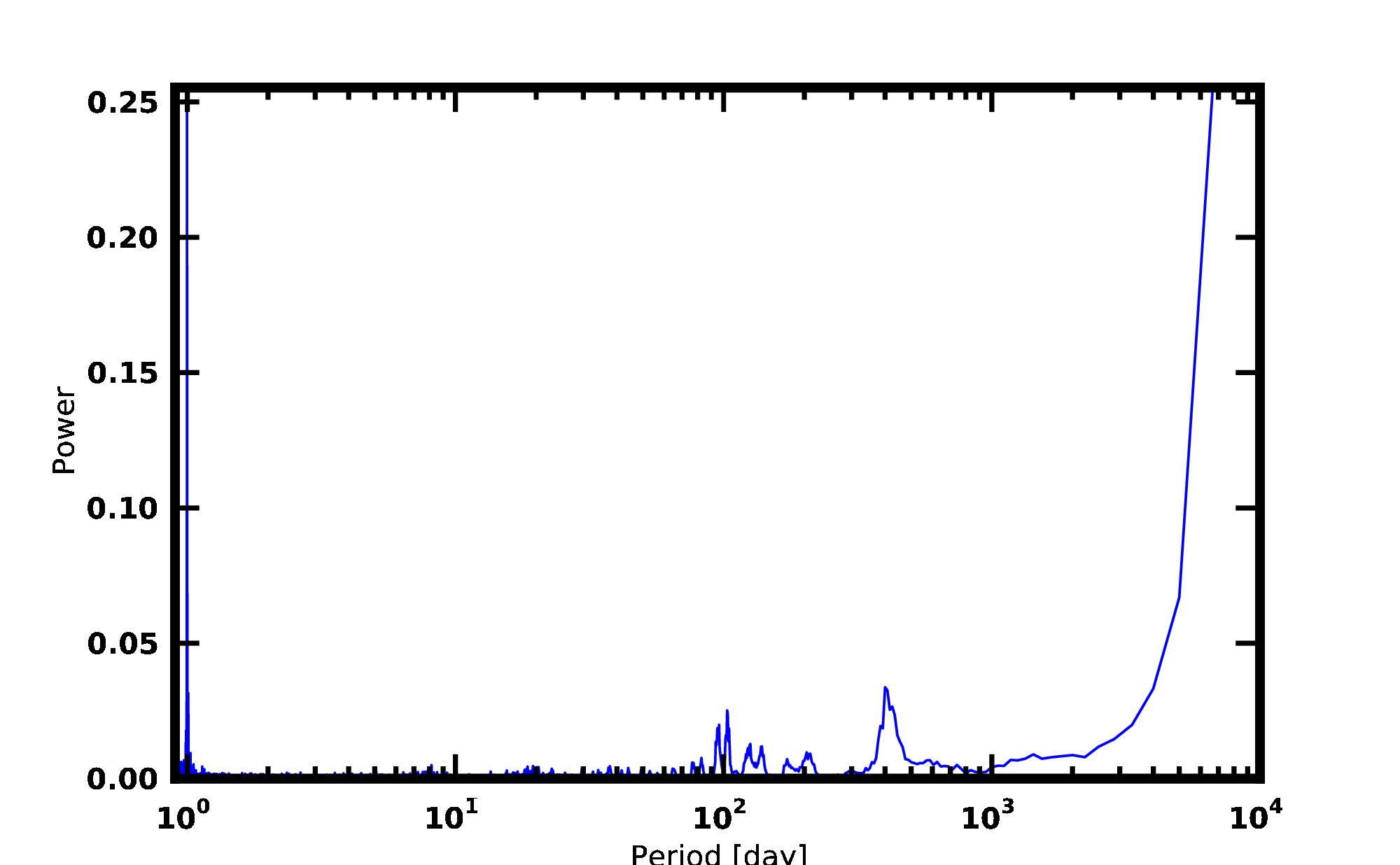} \\
\includegraphics[width=9cm,angle=0]{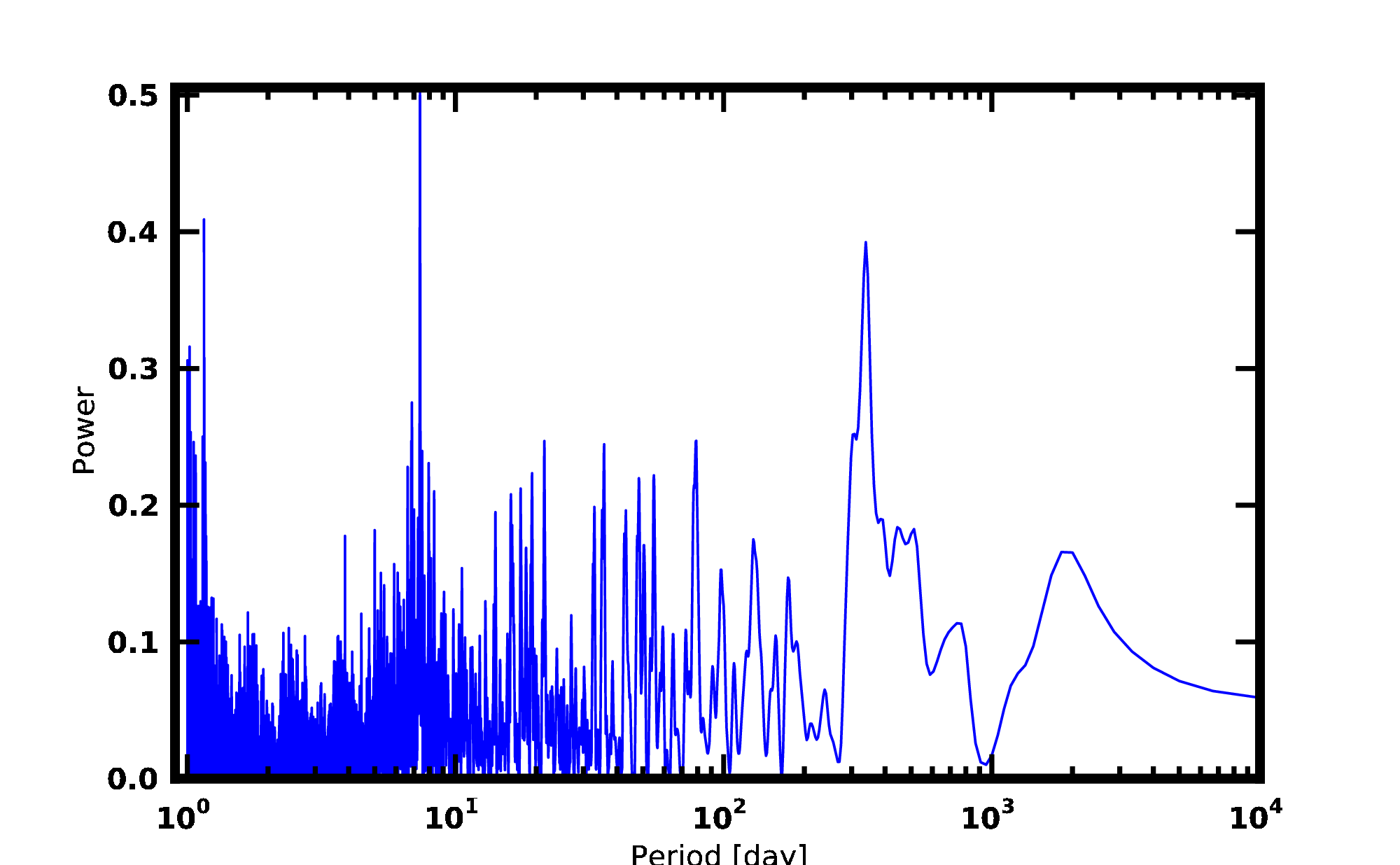} 
\end{tabular}
\caption{
Top : HARPS radial velocity of Gl~433. Middle : window function of the
measurement. Bottom : Periodogram of the HARPS measurements.}
\label{gl433_zero_plan}
\end{figure}

We therefore continued with a classical periodicity analysis, based on
floating-mean periodograms \citep{gilliland1987,
  zechmeister2009}. Fig.~\ref{gl433_zero_plan} (bottom panel) depicts
the periodogram of our RV time series and 
shows a strong power excess around the period 7.3 day. We plot the
window function in Fig.~\ref{gl433_zero_plan} (middle panel) to
identify the typical time sampling but 
found no counter part to this periodicity. We adopted the
normalization of periodograms proposed by \citet{zechmeister2009},
which is such that a power of 1 means that a sine function is a
perfect fit to the data, whereas a power of 0 indicates no improvement
over  a constant model. Hence, the most powerful peak was measured
at a period P$\sim$7.3 days with a power $p_{max}=0.51$. To assess its
fortuity we performed a bootstrap randomization \citep{press1992}: we
generated 10,000 virtual data sets by 
shuffling the actual radial velocities and retaining the dates; for
each set we computed a periodogram; with all periodograms we built a
distribution of power maxima. It appears that, in random data sets,
power maxima exceed $p=0.40$ only once every 100 trials, and never
exceed $p=0.43$ over 10,000 trials. This suggests a False Alarm
Probability (FAP) $<1/10,000$ for the 7.3-d period seen in the
original periodogram. Moreover, for a more precise estimate of this
low FAP value, we made use of \citet{cumming2004} analytic formula :
$FAP=M.(1-p_{max})^{(N-3)/2}$, where $M$ is the number of independent
frequencies in the periodogram, $p_{max}$ its highest power value and
$N$ the number of measurements. We approximated $M$ by 2904/1 (the
ratio between the time span of our observations and the typical 1-day
sampling), and obtained the very low FAP value of $4\times10^{-7}$.

We modeled that first periodic signal by a Keplerian orbit. Even
though the periodicity was securely identified we used $Yorbit$
\citep[an heuristic
algorithm, mixing standard non-linear minimisations and genetic
algorithms; S\'egransan et al. in prep; ][]{bonfils2011} and benefited from a 
global search without {\it a priori}. We converged on a period $P=7.3732 \pm
0.0023$~d, semi-amplitude $K_1=2.99 \pm 0.38$~m/s and eccentricity
$e=0.17\pm0.13$ which, for $M_\star=0.48~{\rm M_\odot}$ converts to $m
\sin i = 5.49\pm0.70~{\rm M_\oplus}$  (Table~\ref{tab_orbit_433}). The rms and 
$\sqrt{\chi^2}$ around the solution decreased to 2.17 m/s and
$2.00\pm0.09$ per degree of freedom, respectively.  

Inspecting the residuals and their periodogram, we measured the power
of the most powerful peak ($P\sim$2900 d; $p_{max}=0.28$) and apply
bootstrap randomization to find it is insignificant ($FAP=22\%$). We
nevertheless found that adding a quadratic drift to the 1-planet model
did improve the solution (rms=1.91 m/s and $\sqrt{\chi^2}=1.79\pm0.09$
per degree of freedom).

\subsection{HARPS+UVES measurements}

\begin{figure}
\begin{tabular}{c}
\includegraphics[width=9cm,angle=0]{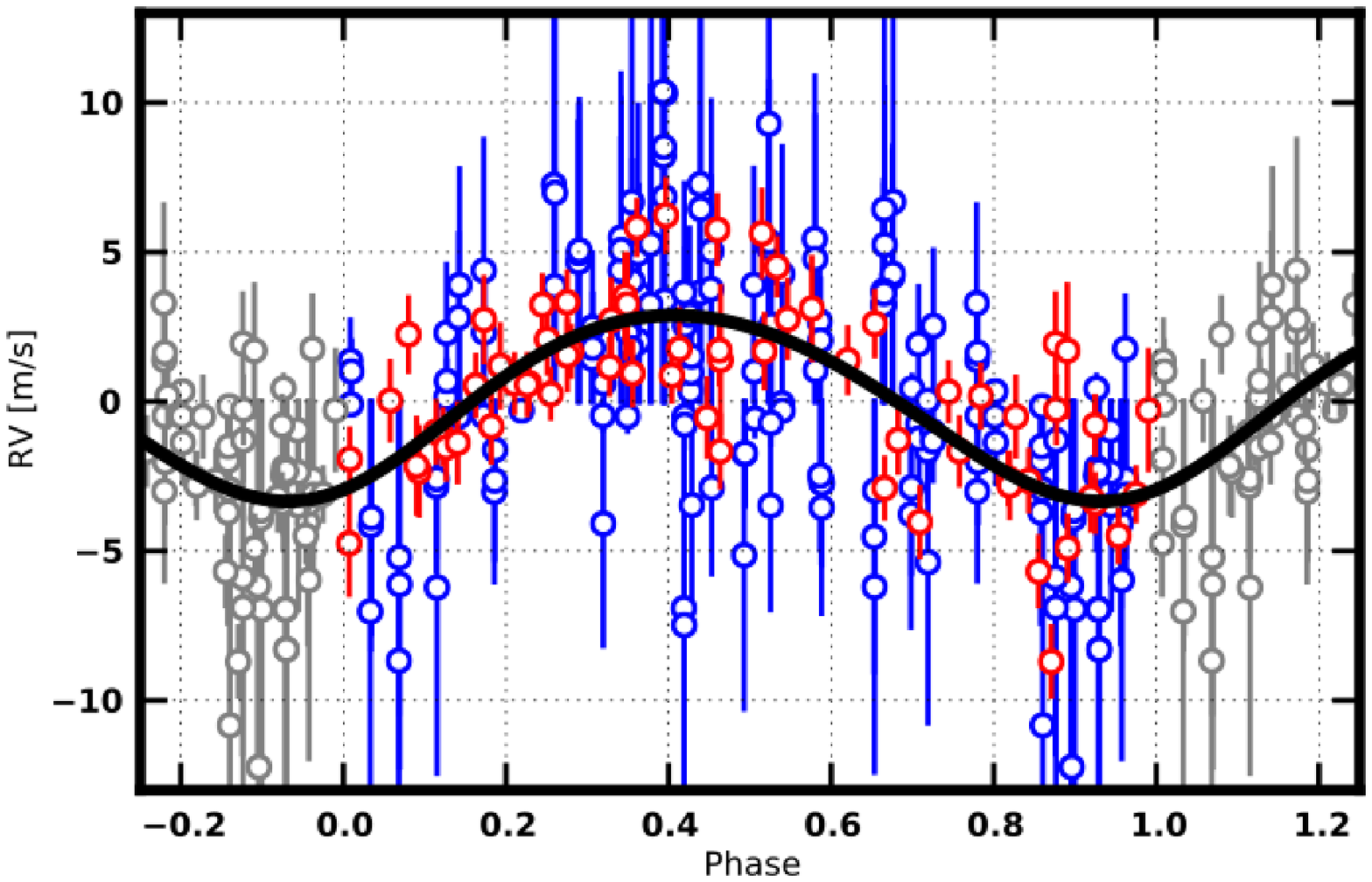} \\
\includegraphics[width=9cm,angle=0]{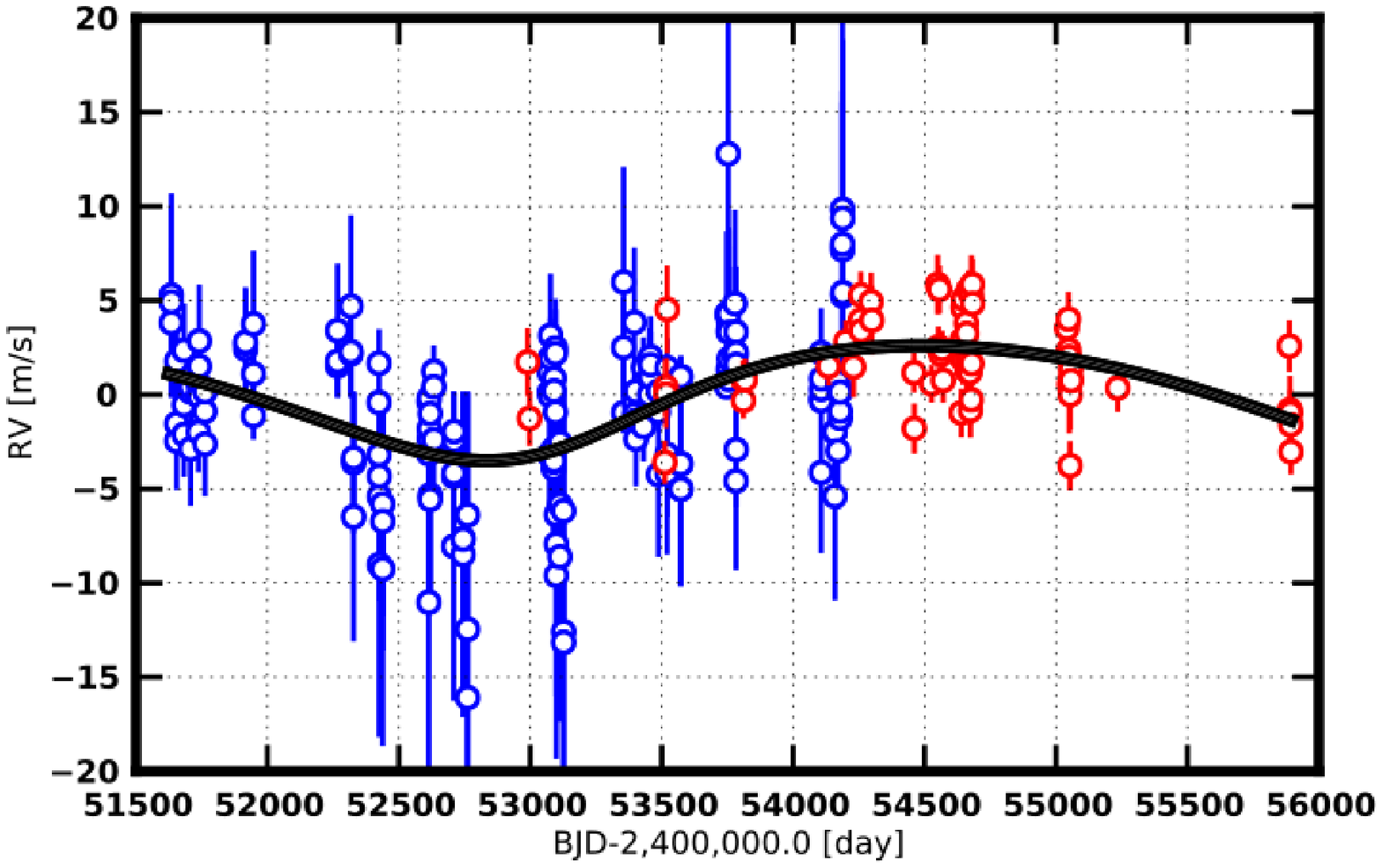} \\
\includegraphics[width=9cm,angle=0]{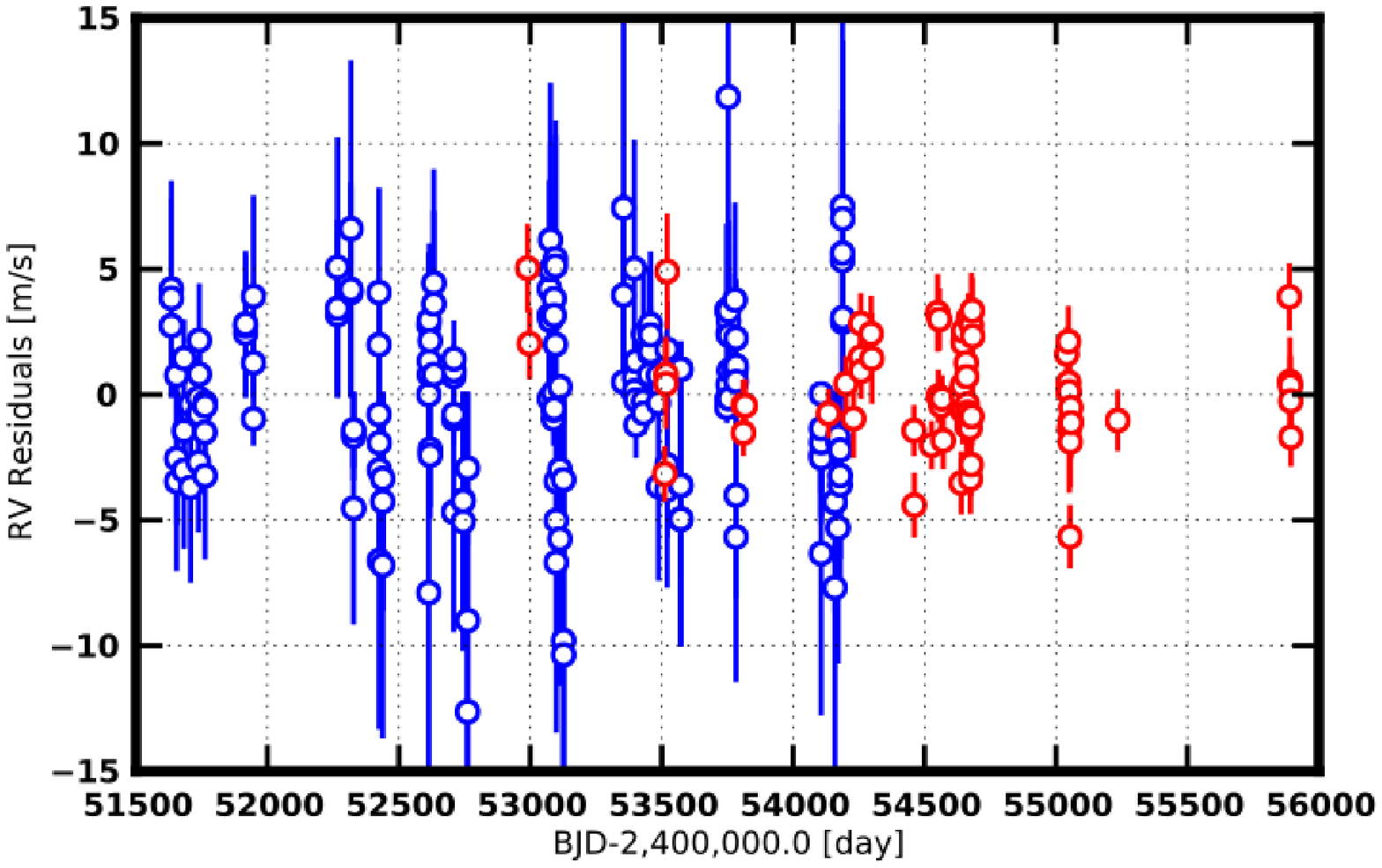} \\
\includegraphics[width=9cm,angle=0]{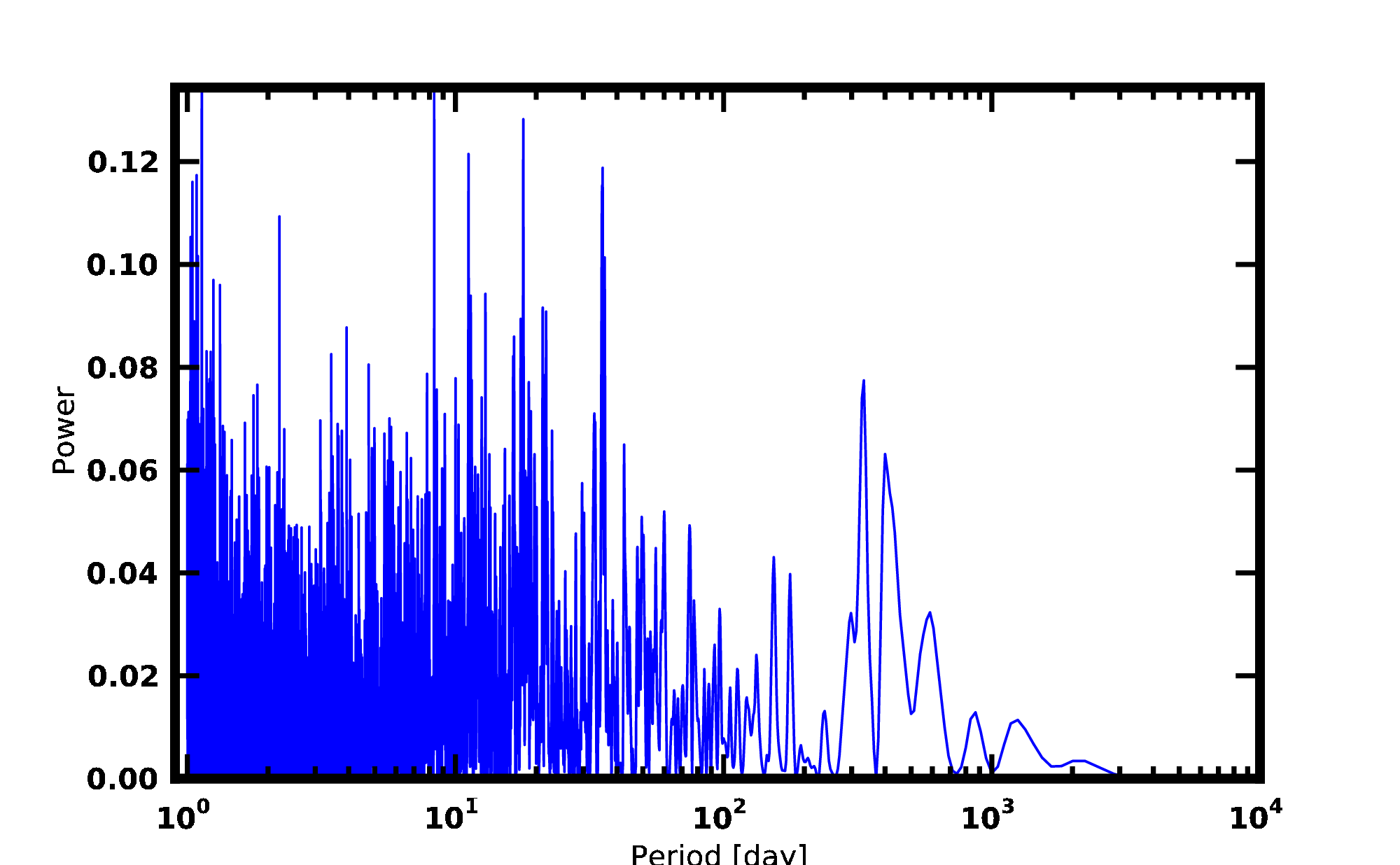}
\end{tabular}
\caption{2-planets model of HARPS (in red) and UVES \citep[in blue,
    from][]{zechmeister2009} radial velocity measurements for Gl~433. 
Top : phased radial velocity for a planet of 7.37~days period. Middle
: radial velocity curve for the 3700~day 
period. Below : residuals and their periodogram.} 
\label{rv_hu}
\end{figure}

Radial velocity measurements of Gl~433 were also obtained with UVES by
\citet{zechmeister2009}. They span 2553 days between March 2000 and
March 2007 and can therefore extend the time span of our observations
by more than 3 years (see Fig.~\ref{rv_hu} - second panel).  

Hence we pooled together UVES and HARPS data and performed again a
{\it 1 planet} fit with $Yorbit$. We converged on the same solution
with more precise orbital parameters ($P=7.37131\pm0.00096$~d;
$K_1=2.846\pm0.248$~m/s; $e=0.13\pm0.09$). The residuals around the
solution show a power excess at a similar period than before
($P\sim$2857 d; $p_{max}=0.19$) that now appears more
significant. With 10,000 trials of bootstrap randomization we found a
$FAP\sim0.2\%$. Although such signal requires additional measurements
for confirmation it justified the addition of a second planet to our
model.

With a model composed of 2 planets on Keplerian orbits, $Yorbit$
converges on periods of 7.37029$\pm$0.00084 and 3690$\pm$250 days,
with semi-amplitudes of 3.113$\pm$0.224 and 3.056$\pm$0.433 m/s and
eccentricities 0.08277$\pm$0.07509 and 0.17007$\pm$0.09428
(see. Table~\ref{tab_orbit_433_hu} for the full set of parameters). The
solution has a rms=2.39 m/s
and a $\sqrt{\chi^2}=1.30\pm0.05$ per degree of freedom, showing that
the combination of HARPS and UVES data sets largely improved the
parameter uncertainties.

\subsection{The planetary system around Gl~433}

In the past, Gl~433 was announced as hosting a brown dwarf of 30
Jupiter mass with an orbital period of $\sim$500~days from astrometric
measurements \citet{bernstein1997}. But all radial velocity
measurments published after, as the present one, reject this
detection.

Apparent Doppler shifts may also originate from stellar surface
inhomogeneities, such as plages and spots, which can break the balance
between light emitted in the red-shifted and the blue-shifted
parts of a rotating star
\citep[e.g.][]{saar1997,queloz2001,desort2007}. However the large rotation  
period ($>$40~days) of Gl~433 ensures that the observed doppler shift
of 7.37~days does not originate from the stellar activity. A
search of correlation between the 7.37~days radial velocity period and
H$_{\alpha}$, CaII-index or bissector is
unsuccessful in our HARPS data. Such signal is easily detected when
the radial velocity variation is due to activity for a period of
$\sim$35~days (and {\it a fortiori} for all shorter periods) \citep[see the
  case of Gl~176 and Gl~674]{bonfils2007,forveille2009}. This is a
strong evidence that the $\sim$7~d period is not due to stellar surface
inhomogeneities but to the presence of a planet orbiting around Gl~433.

This planet, Gl~433b, belongs to the category of
super-Earth with a minimum mass of $\sim$5.8~M$_{\oplus}$. At a separation
of 0.058AU from its star, Gl~433b is illuminated by a bolometric flux,
per surface unity, 10 times higher than what the Earth receives.

The long period variability detected in the HARPS and UVES data set
may have several origins. The signal can have as origin a
$\sim$10-year period planet of 50~M$_{\oplus}$. A possible origin may 
also be an effect of a long term stellar magnetic cycle (like the
so-called 11-year solar cycle) during which the fraction of the
stellar disk covered by plages varies. In such region the magnetic
field attenuates the convective flux of the blueshifted plasma and
impacts the mean observed radial velocity in the order of few~m/s or
more \citep{meunier2010}. \citet{gomes2011} do detect large period
variation for numerous activity proxy (CaII, NaI, HeI, H$_{\alpha}$
lines) of Gl~433 coherent with a period of $\sim$10~years. The RV
variability at long period is well correlated with these
activity proxies (Gomes da Silva et al. in prep). Therefore, we
favor a Magnetic Cycle as the origin of the long term signal.

\begin{table}
\caption{
Fitted orbital solution for the HARPS data of Gl~433}
\begin{tabular}{ll} \hline
 & Gl~433b  \\ \hline \hline
$P$ [days]                  & 7.373$\pm$0.002\\
$e$                         & 0.17$\pm$0.13\\
$T0$ [JD - 2400000]         &  54597.0$\pm$1.0\\
$\omega$ [deg]                  & 136$\pm$47 \\
$K_1$ [m/s]                 & 2.9$\pm$0.4\\
%$f(m)$ [$10^{13}$M$_{\odot}$] & 0.252 \\ 
%$a_1.\sin{i}$ [$10^{5}$AU]   & 0.21\\
$M_2.\sin{i}$ [M$_{\oplus}$]  & 5.49 \\
$a$ [AU]                    & 0.058 \\ \hline
$N_{meas}$                   & 67  \\
$Span$ [days]               & 2904 \\
r.m.s [m/s]         & 2.17 \\
$\sqrt{\chi^2}$             & 2.00 \\
\end{tabular}
\label{tab_orbit_433}
\end{table}

\begin{figure}
\begin{tabular}{c}
\includegraphics[width=9cm,angle=0]{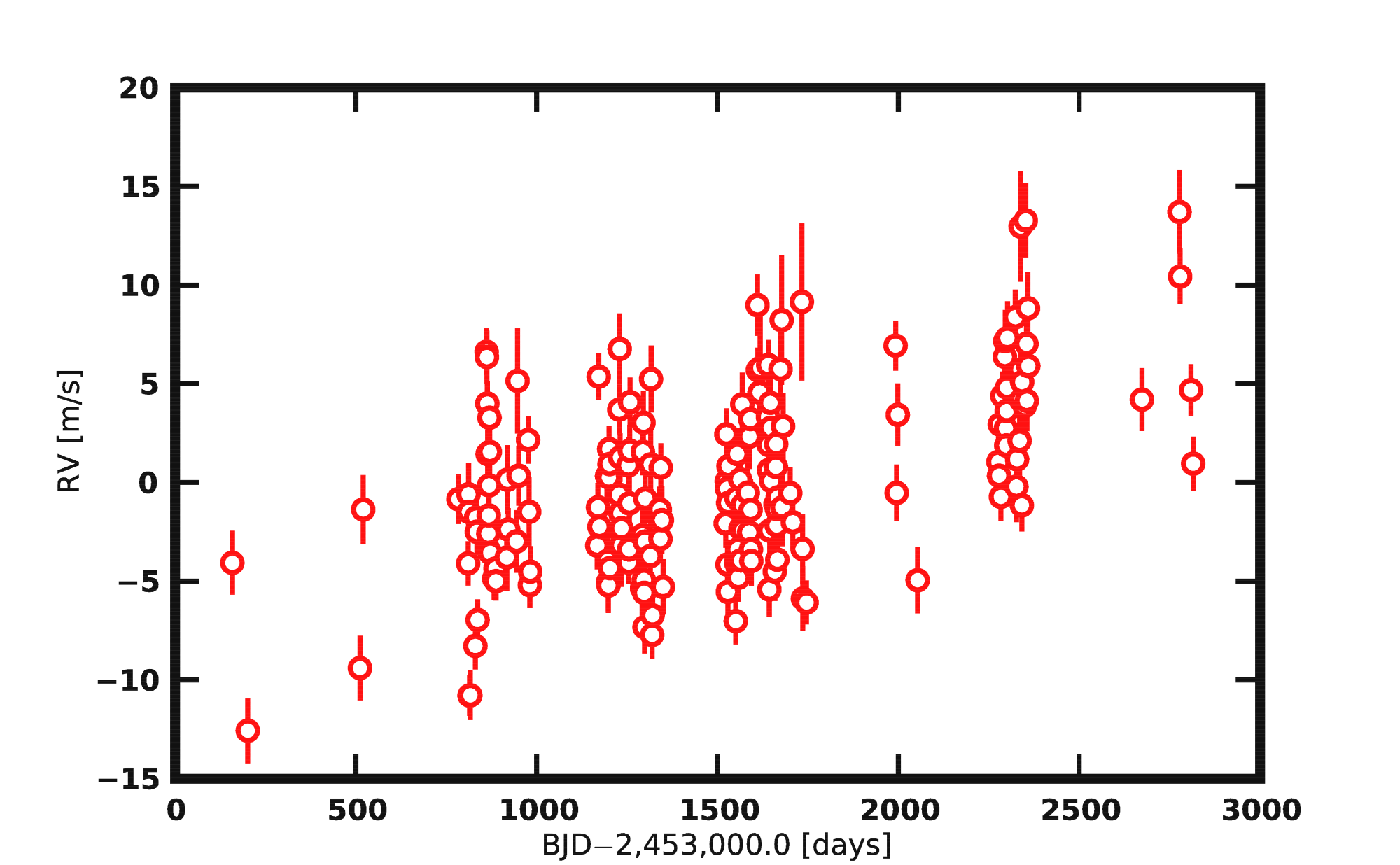} \\
\includegraphics[width=9cm,angle=0]{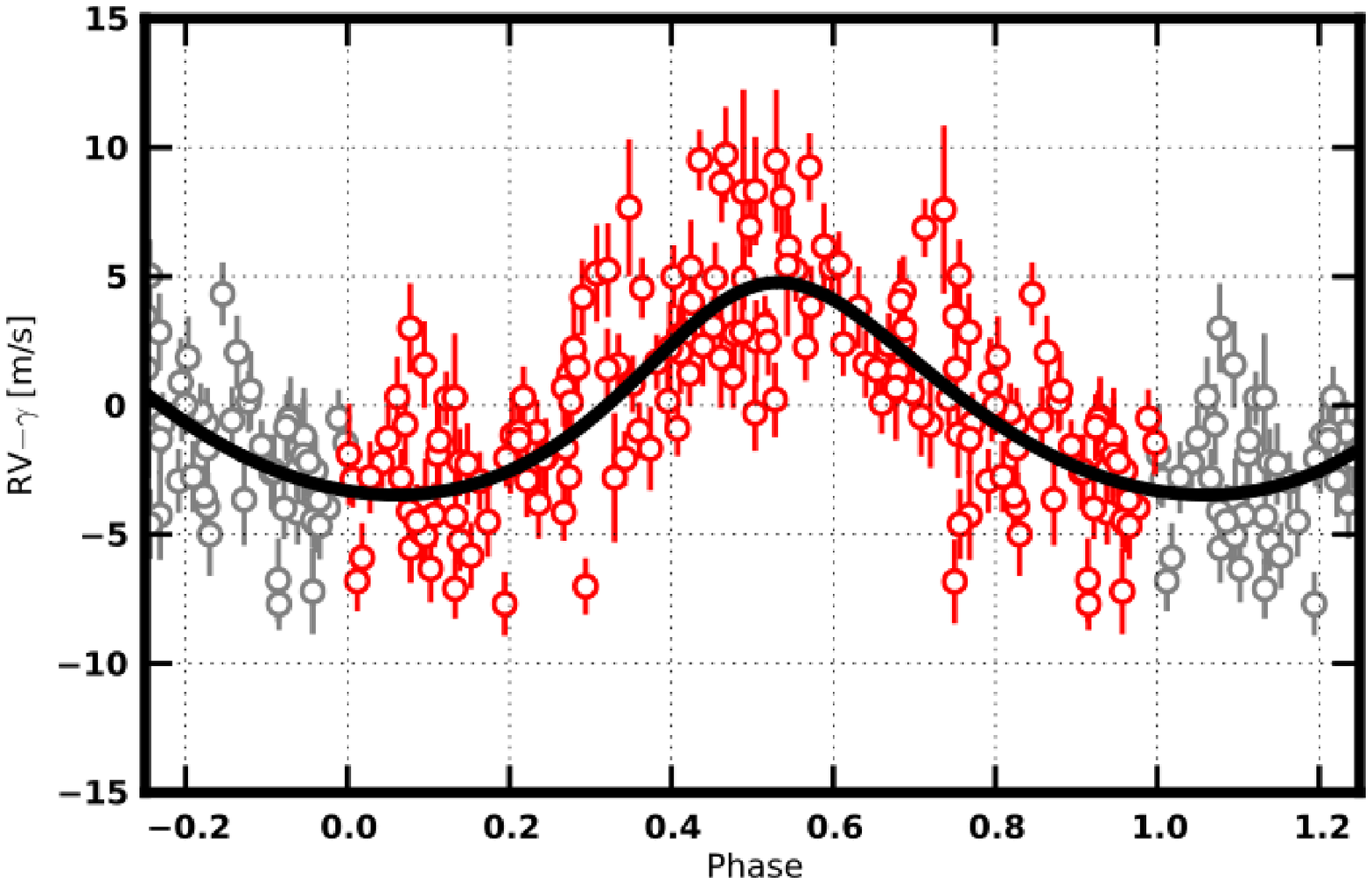} 
\end{tabular}
\caption{
Top : HARPS radial velocity of Gl~667C. Bottom : one planet plus a
linear drift model.} 
\label{gl667_zero_one_plan}
\end{figure}

\begin{table}
\caption{
Fitted orbital solution for the HARPS and UVES data of Gl~433}
\begin{tabular}{lll} \hline
 & Gl~433b & Gl~433c \\ \hline \hline
$P$ [days]                  & 7.3709$\pm$0.0008 & 3693$\pm$253 \\ 
$e$                         & 0.08$\pm$0.08     & 0.17$\pm$0.09\\
$T0$ [JD - 2400000]         & 54287$\pm$1 & 56740$\pm$462\\
$\omega$ [deg]                  & -156$\pm$54 & -154$\pm$36 \\
$K_1$ [m/s]                 & 3.11$\pm$0.23      & 3.1$\pm$0.5 \\
%$f(m)$ [$10^{13}$M$_{\odot}$] & 0.225            & 139.6 \\ 
%$a_1.\sin{i}$ [$10^{5}$AU]   & 0.20             & 10.9  \\
$M_2.\sin{i}$ [M$_{\oplus}$]  & 5.79             & 44.6   \\
$a$ [AU]                    & 0.058            & 3.6 \\ \hline
$N_{meas}$                   & \multicolumn{2}{c}{233}  \\
$Span$ [days]               & \multicolumn{2}{c}{4259} \\
r.m.s. [m/s]                & \multicolumn{2}{c}{2.39} \\
$\sqrt{\chi^2}$             & \multicolumn{2}{c}{1.30} \\
\end{tabular}
\label{tab_orbit_433_hu}
\end{table}

\section{Orbital analysis of Gl~667C}

\begin{figure*}
\begin{tabular}{cc}
(a) \includegraphics[width=8cm,angle=0]{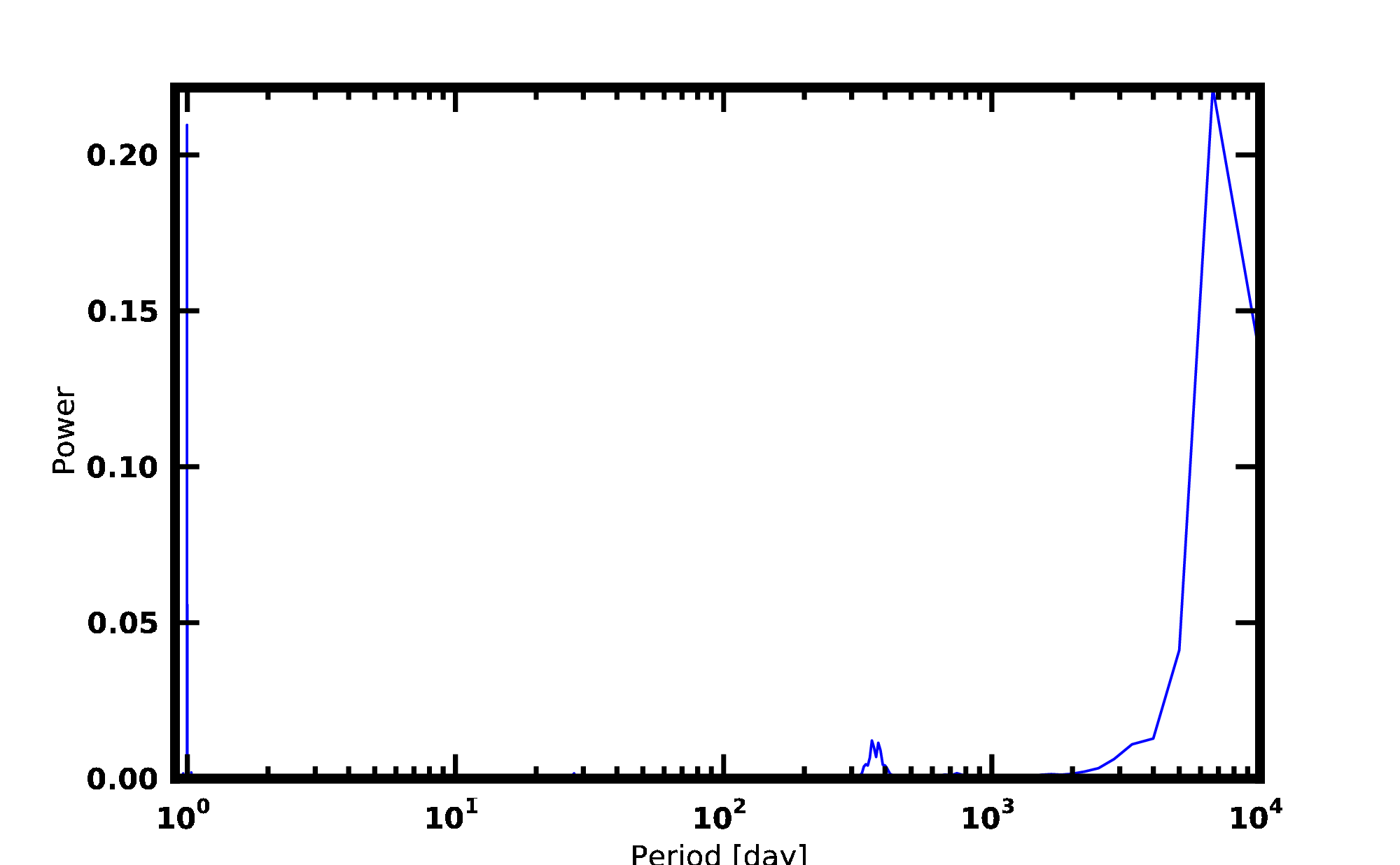} &
(b) \includegraphics[width=8cm,angle=0]{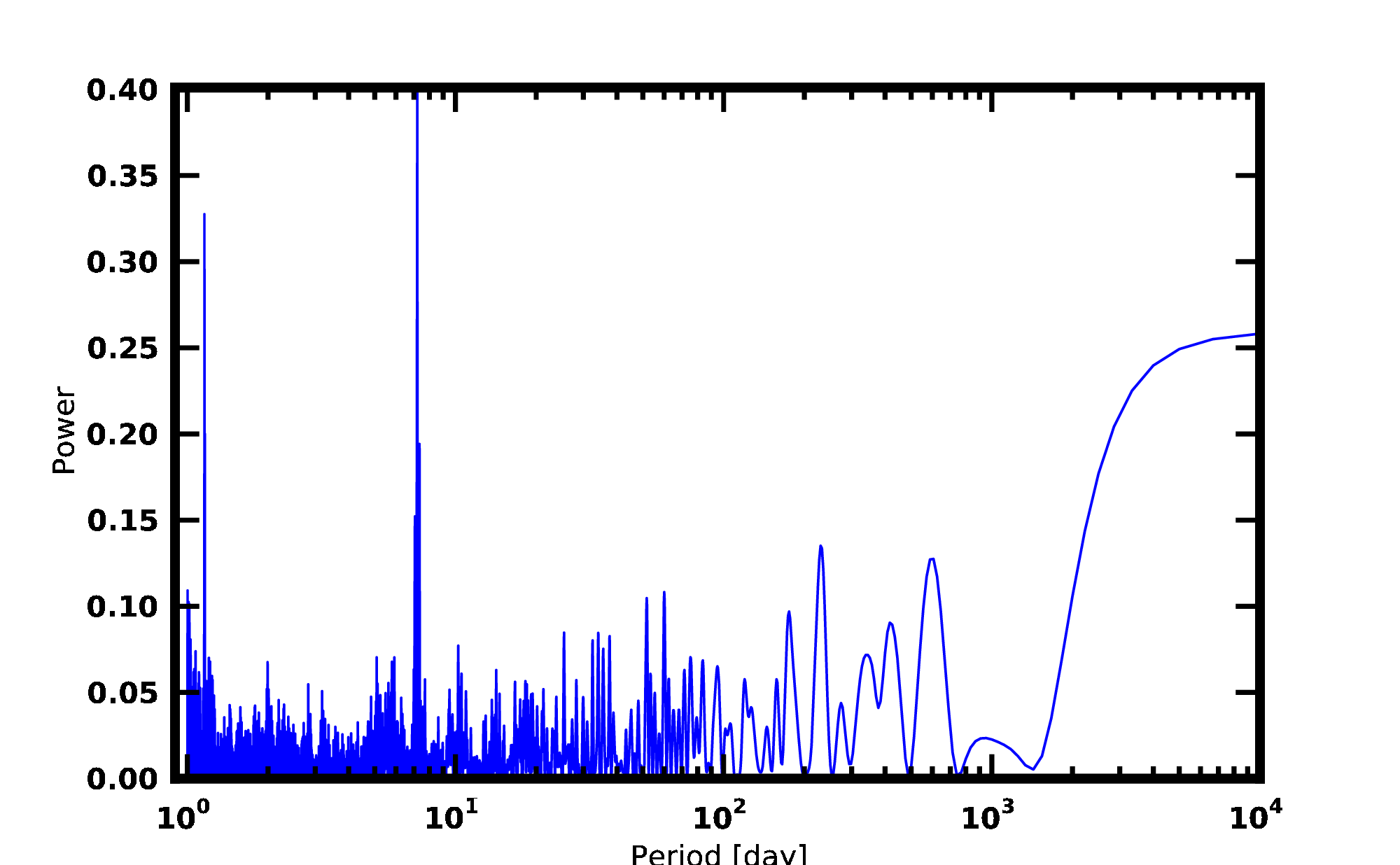} \\
(c) \includegraphics[width=8cm,angle=0]{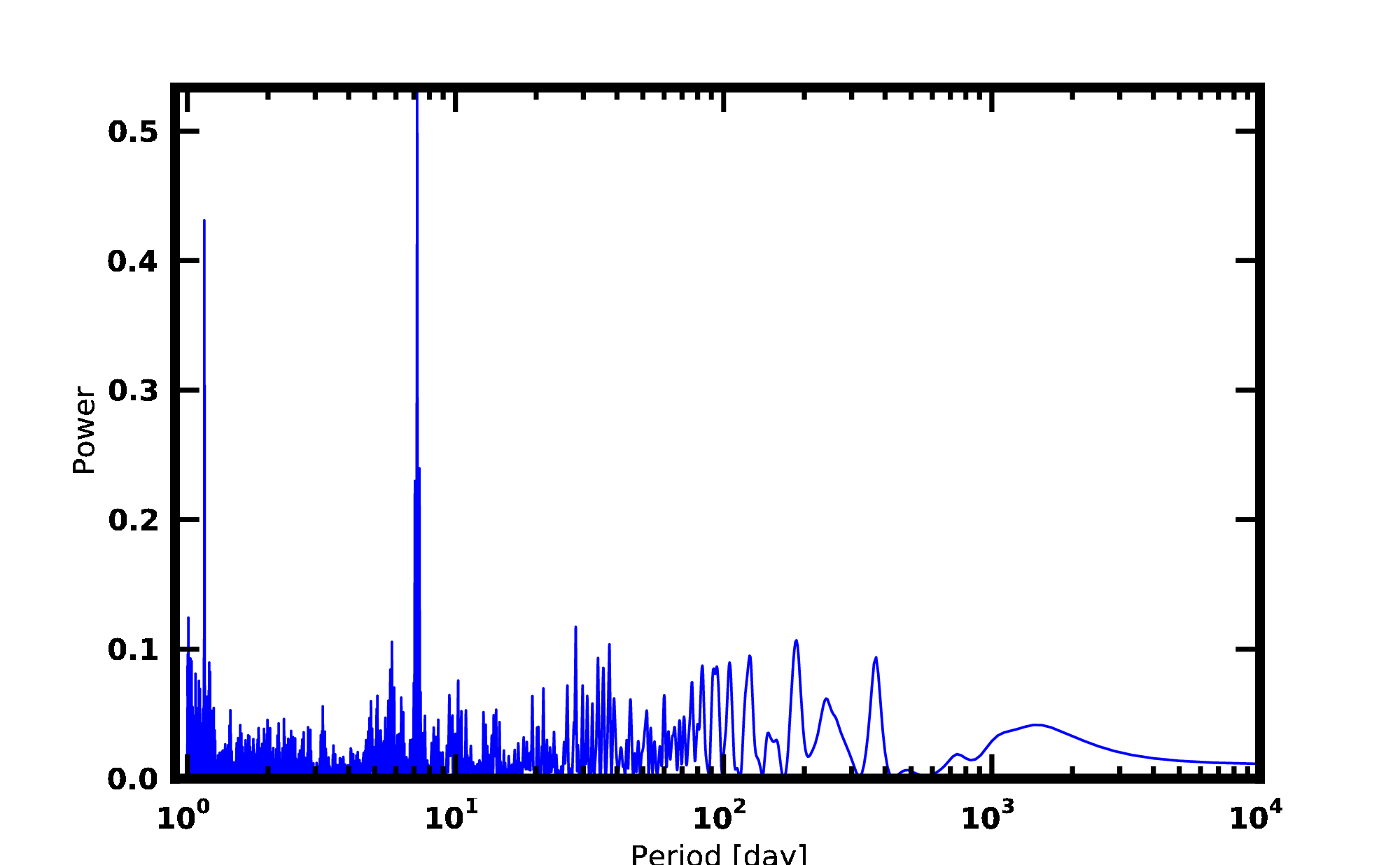} &
(d) \includegraphics[width=8cm,angle=0]{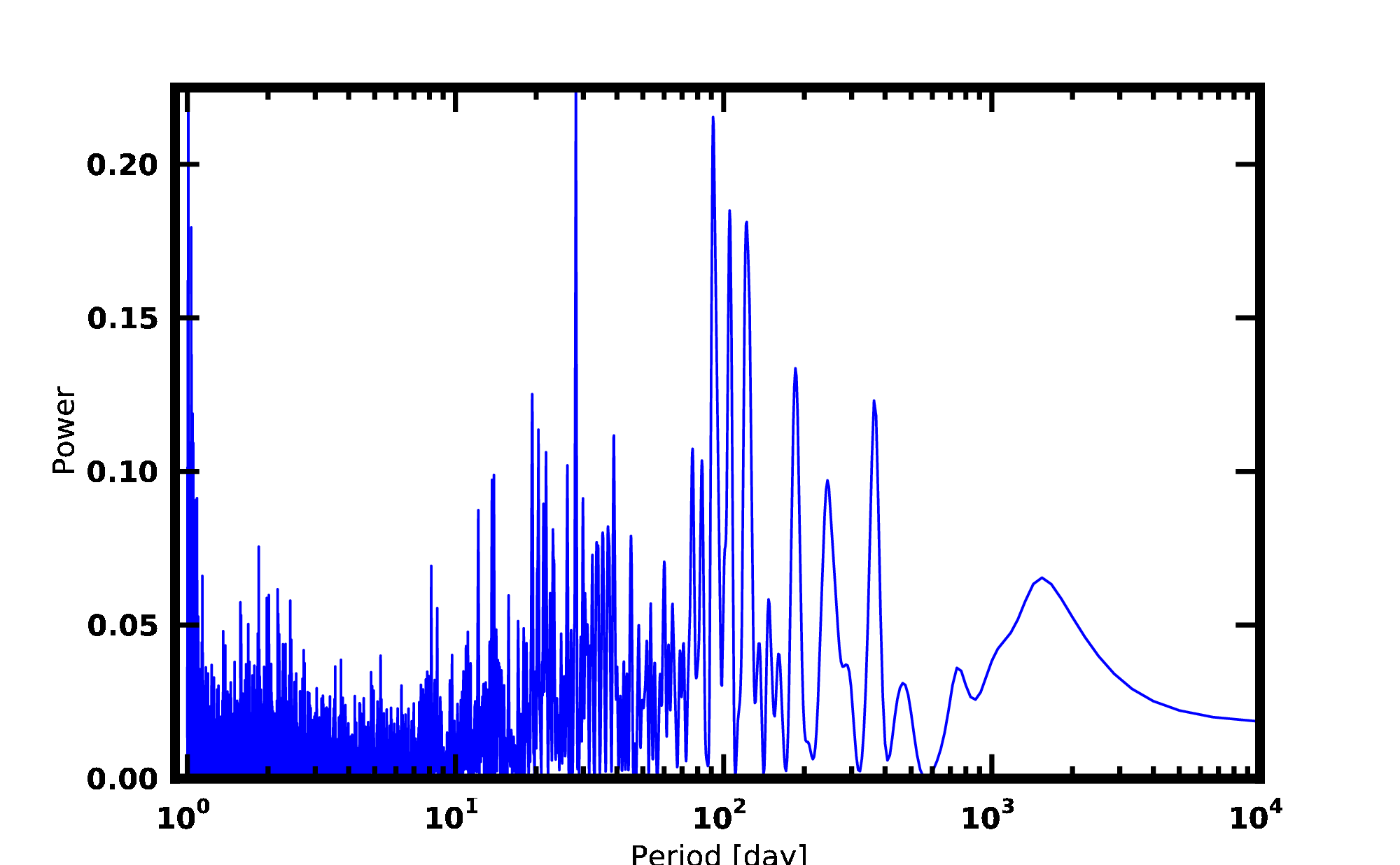}
\end{tabular}
\caption{
From the top to the bottom : (a) Window function of the RV
measurement; (b)periodograms of the Gl~667C RV measurements; (c)
periodograms of the Gl~667C RV measurements after a drift removal; (d)
periodograms of the residual after subtraction of a drift + 1 planet
model.} 
\label{gl667_perio_1plan}
\end{figure*}

We obtained 179 measurements of Gl~667C radial velocity  spanning
2657 days between June 2004 and September 2011. The data adds 36 points
to those analyzed in \citet{bonfils2011} and extend their time span
by 1070 days. They have
uncertainties of $\sim$1.3 m/s (dominated by photon noise) and both
rms and $\sqrt{\chi2}$ values (resp. 4.3 m/s and 3.04 per degree of
freedom) indicate a variability above those uncertainties. All RVs are
shown as a function of time in Fig.~\ref{gl667_zero_one_plan} (top
panel) whereas their window function and floating-mean periodogram are 
presented Fig.~\ref{gl667_perio_1plan} (top two panels).

\subsection{A one planet and linear fit solution}

The long-term drift seen in Gl~667C RVs agrees well with the
line-of-sight acceleration induced by its companion stellar pair
Gl~667AB which is about $GM_{AB}/r^2_{AB-C} \sim
3$m/s/yr (for a total mass $M_{AB}$ of 1.27M$_{\odot}$ and a
separation of $\sim$300 AU). Even without removing that linear drift
the periodogram
additionally shows a peak at P$\sim$7.2 days (see
Fig.~\ref{gl667_perio_1plan}, panel b). We do remove an adjusted
drift ($\dot{\gamma} = 1.59 \pm 0.14$ m/s/yr) to make the periodic signal
even stronger ($p_{\rm max} = 0.53$; Fig.~\ref{gl667_perio_1plan}
panel c). To measure the FAP of the 
7.2~days signal we ran 10,000
bootstrap randomization and found no power stronger than 0.18
suggesting a FAP$<<1/10,000$. We also computed the FAP with
\citep{cumming2004}'s prescription: $FAP = M.(1-p_{\rm
  max})^{(N-5)/2}$. We approximate $M$ by $2657/1$ (the ratio between
the time span and the typical sampling of our observations) and
obtained the extremely low FAP value of $\sim10^{-25}$. Note that, on
Fig.~\ref{tab_gl667_perio_1plan} (panel c),
a significant and much less powerful peak is seen around P=1.0094 day
and corresponds to an alias of the 7.2-day peak with the typical
$\sim$1~day sampling.

We pursued by adjusting with $Yorbit$ the RVs to a model composed of 1
planet plus a 
linear drift and converged robustly on the orbital elements reported in
Table.~\ref{gl667_perio_1plan}. This model reduced the 
rms and $\sqrt{\chi2}$ per degree of freedom to 2.51 m/s and 2.02,
respectively. They nevertheless remain above the photon noise and
instrumental uncertainties, what prompted us to continue the analysis
with the residuals and try more complex models.

\begin{table}
\caption{1 planet + 1 linear drift orbital solution for Gl~667}
\begin{tabular}{lllll} \hline
 & Gl~667Cb &  &  & \\ \hline
$P$ [days]                  & 7.1989$\pm$0.0015 & & & \\
$e$                         & 0.11$\pm$0.07 & & & \\
$T0$ [JD - 2400000]         & 54373.4$\pm$0.7 & & & \\
$\omega$ [deg]              & -22.$\pm$36. & & & \\
$K_1$ [m/s]                 & 3.9$\pm$0.3 & & & \\
%$f(m)$ [$10^{13}$M$_{\odot}$] & 0.458 & & & \\ 
%$a_1.\sin{i}$ [$10^{5}$AU]   & 0.26103 & & & \\
$M_2.\sin{i}$ [M$_{\oplus}$]  & 5.69 & & &  \\
$a$ [AU]                    & 0.0504  & & & \\ \hline
$\gamma$ [km/s]             & 6.55$\pm$0.02       & & & \\
$\dot{\gamma}$ [m/s/yr]            & 1.26$\pm$0.18  & & & \\
r.m.s [m/s]         & 2.51  & & & \\
$\sqrt{\chi^2}$             & 2.02  & & & \\ \hline \hline
\end{tabular}
\label{tab_gl667_perio_1plan}
\end{table}

\begin{figure*}
\begin{tabular}{cc}
(a) \includegraphics[width=8cm,angle=0]{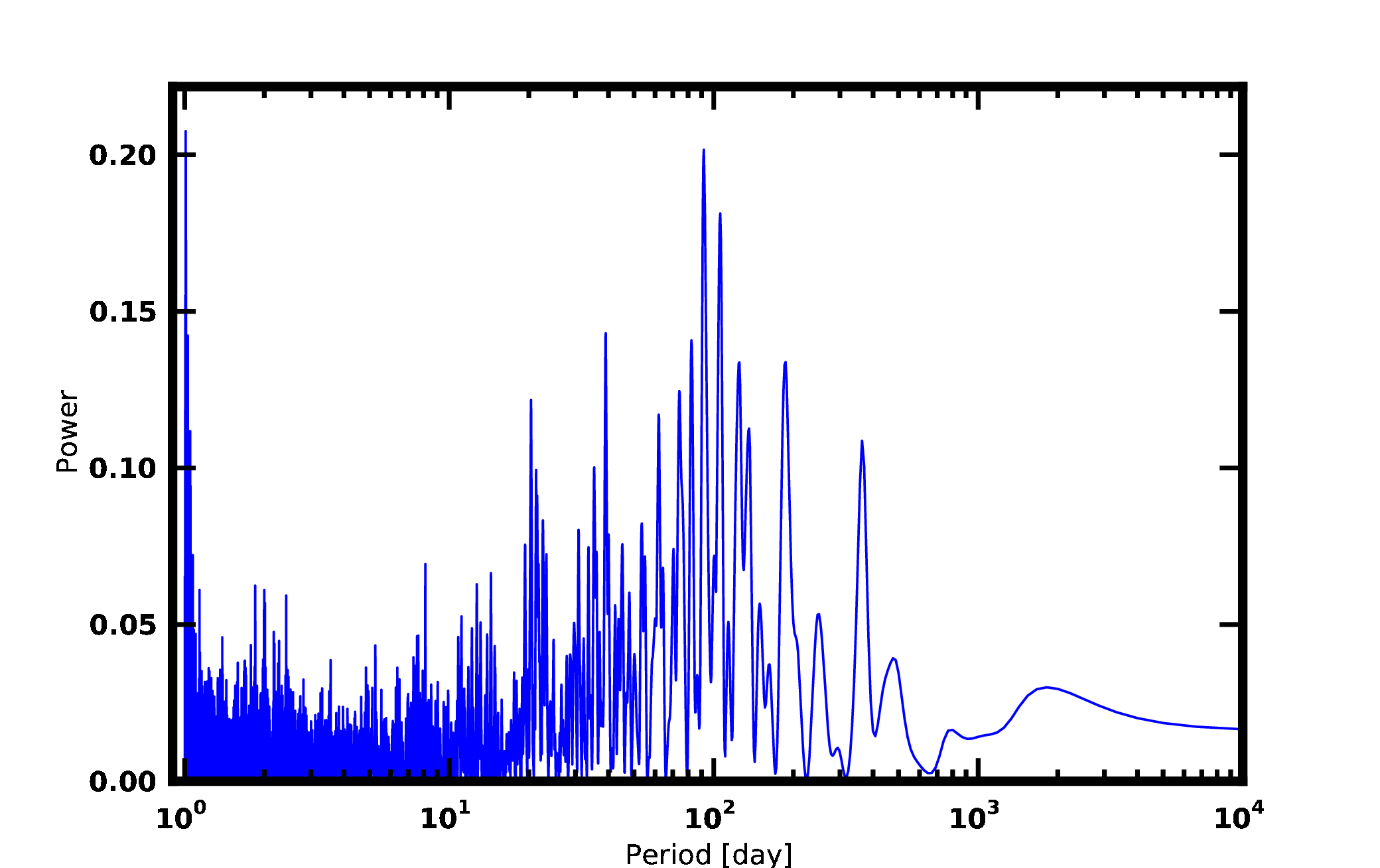} &
(b) \includegraphics[width=8cm,angle=0]{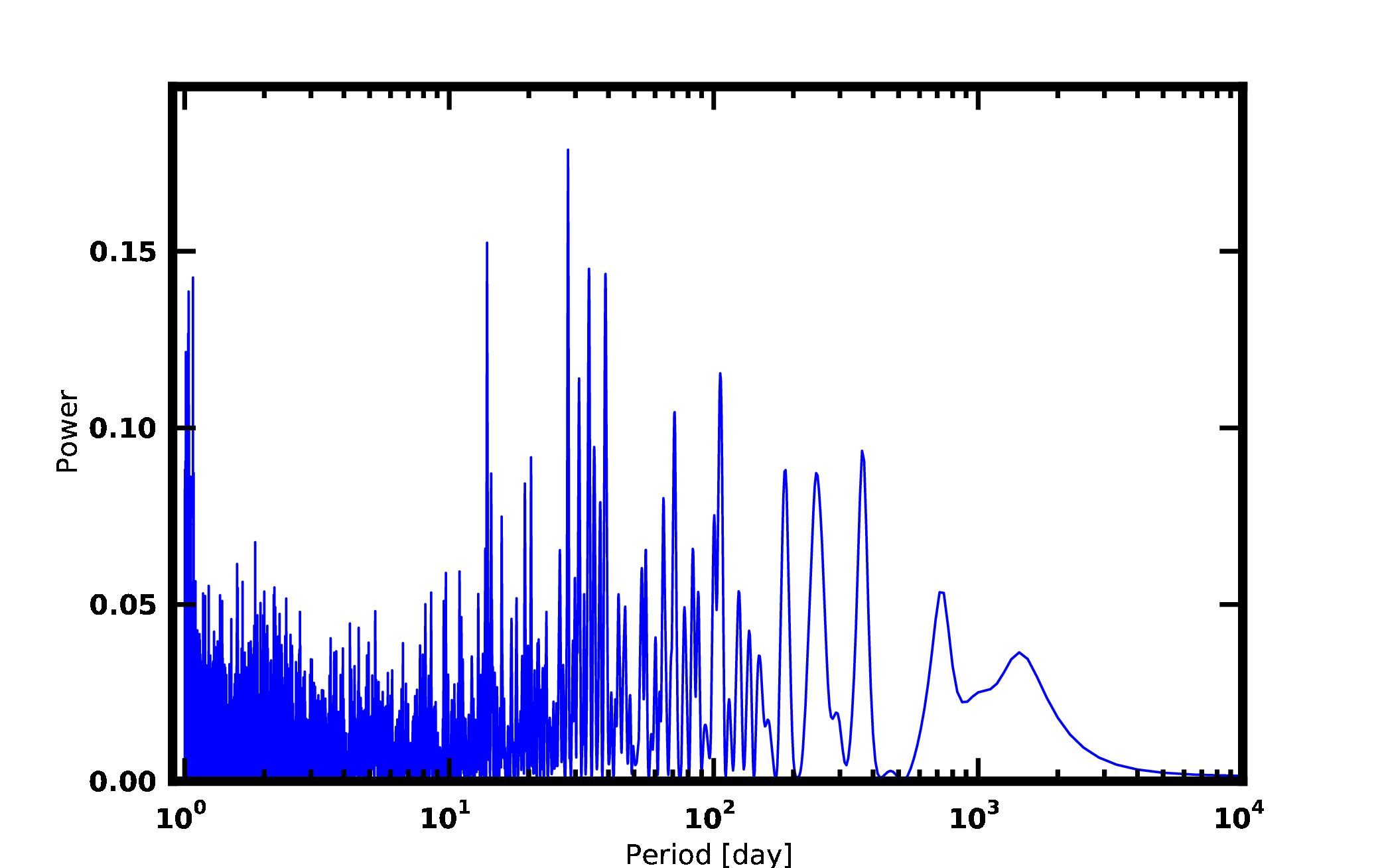} \\
(c) \includegraphics[width=8cm,angle=0]{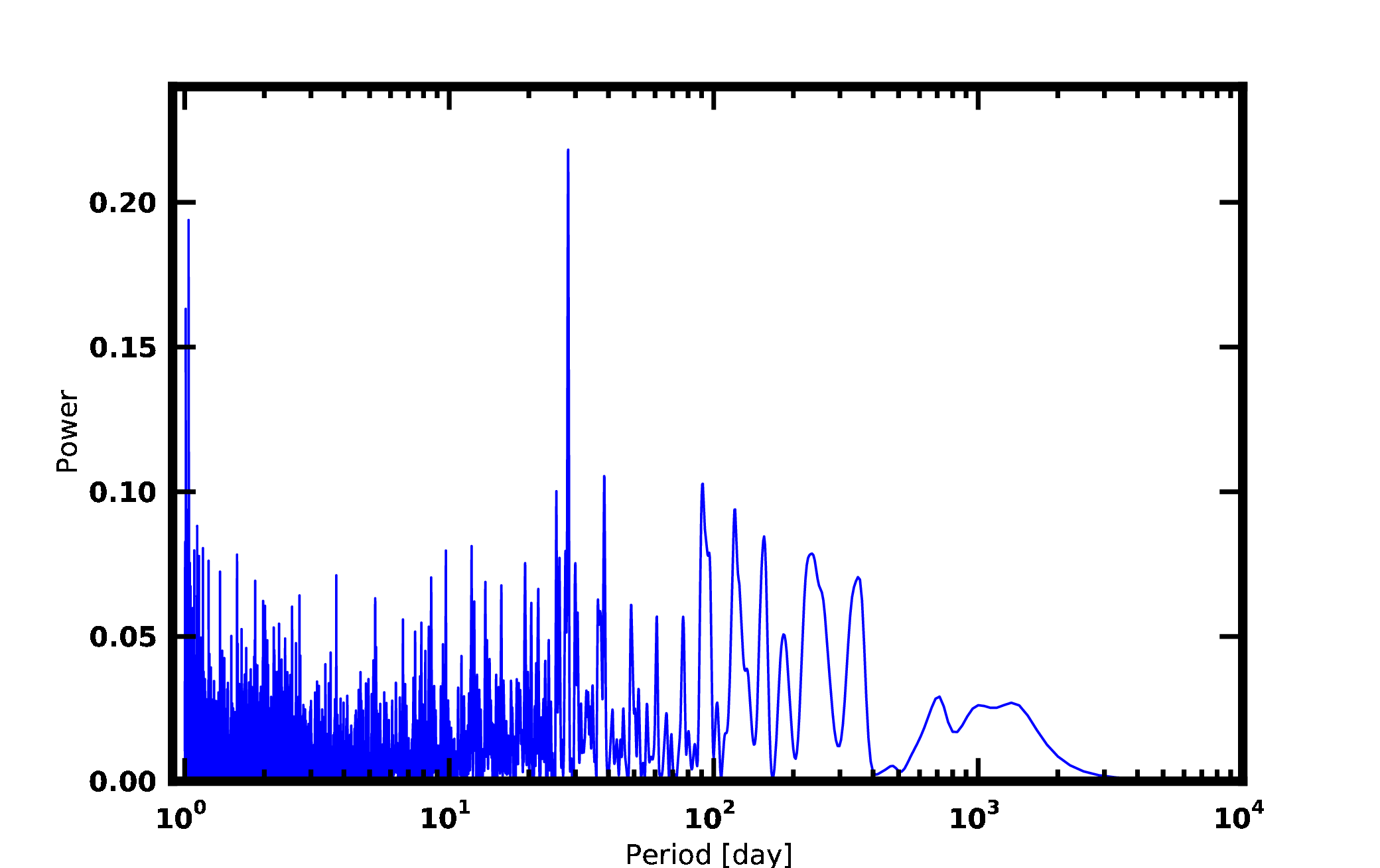} &
(d) \includegraphics[width=8cm,angle=0]{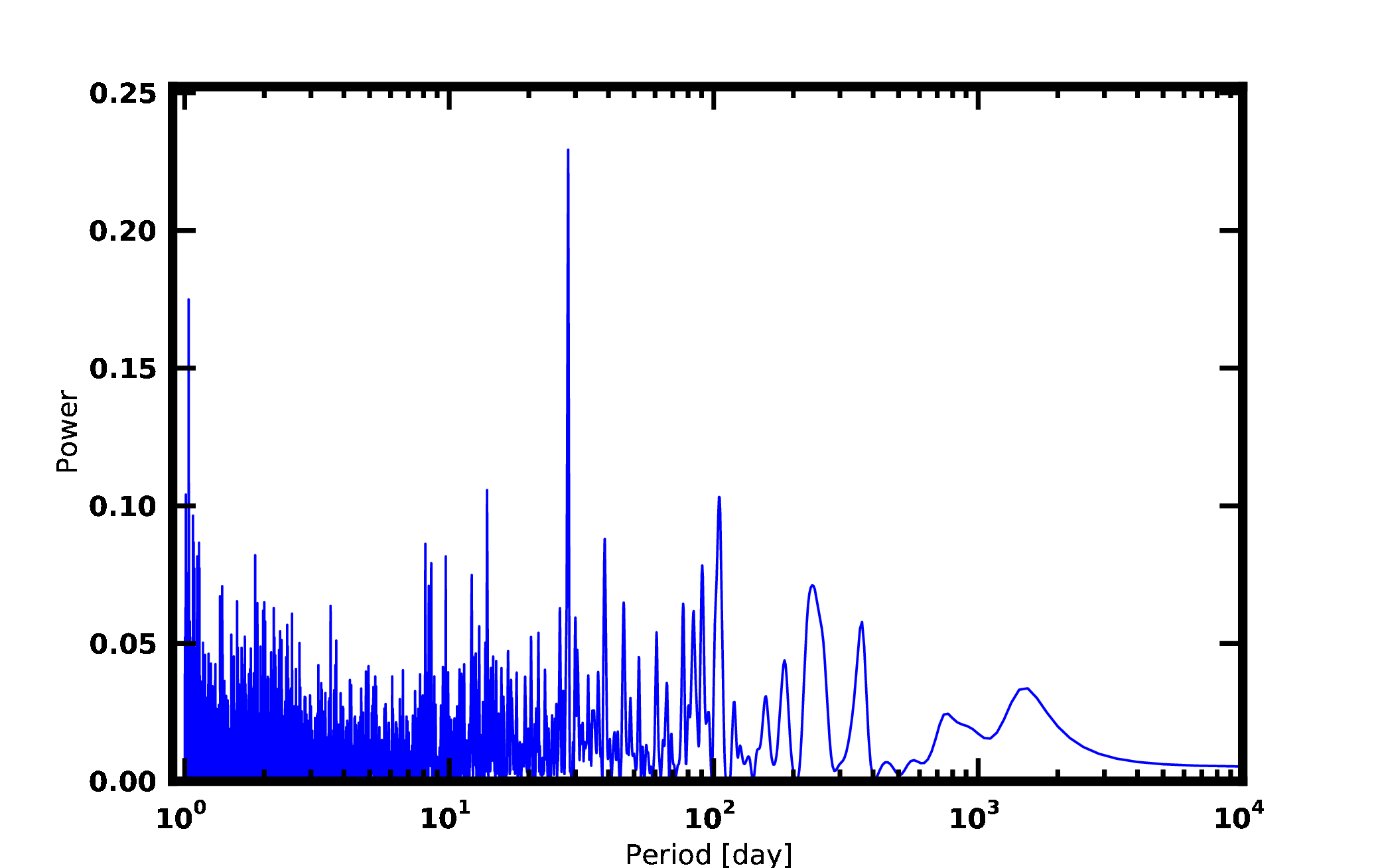} \\
(e) \includegraphics[width=8cm,angle=0]{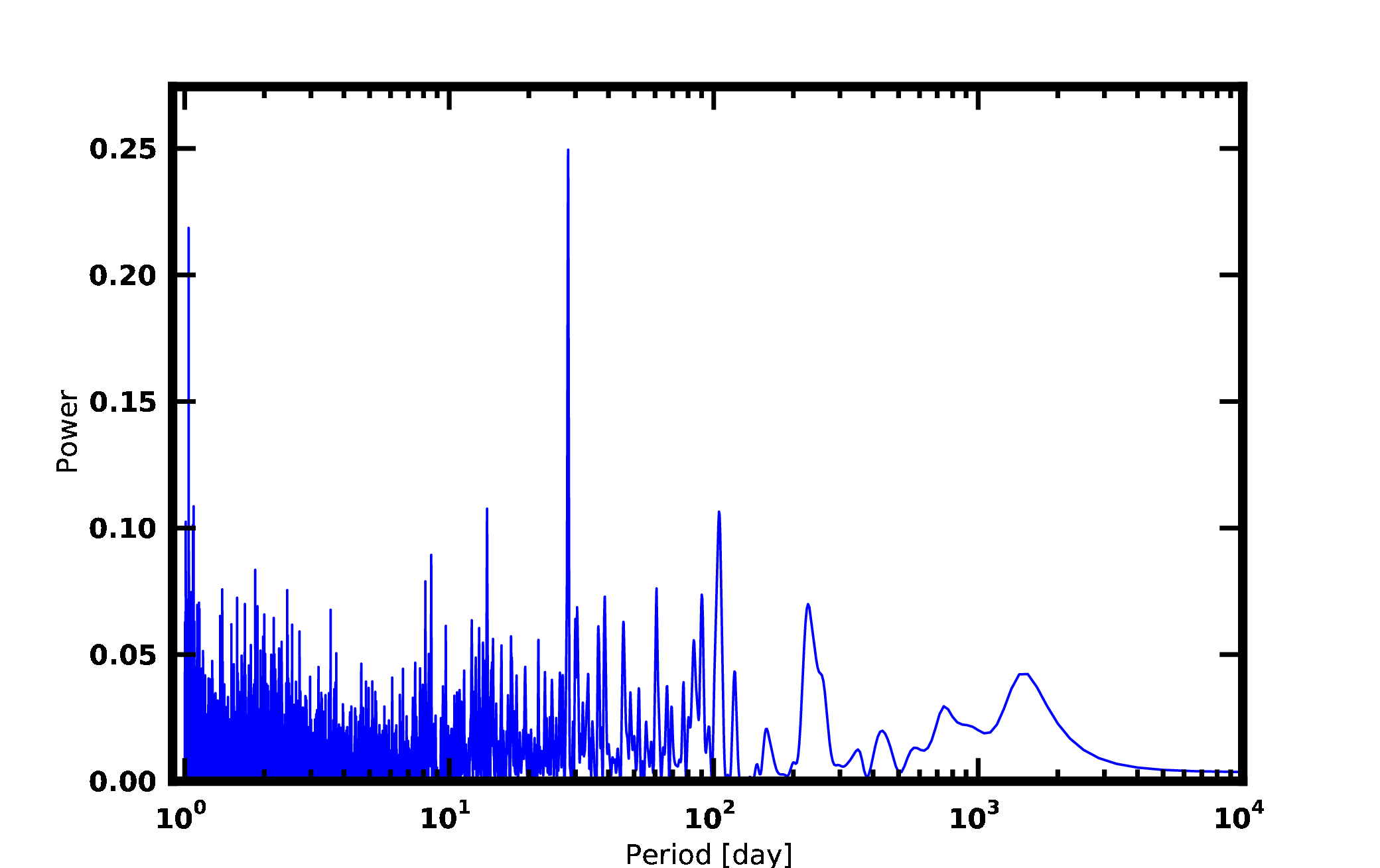} &
(f) \includegraphics[width=8cm,angle=0]{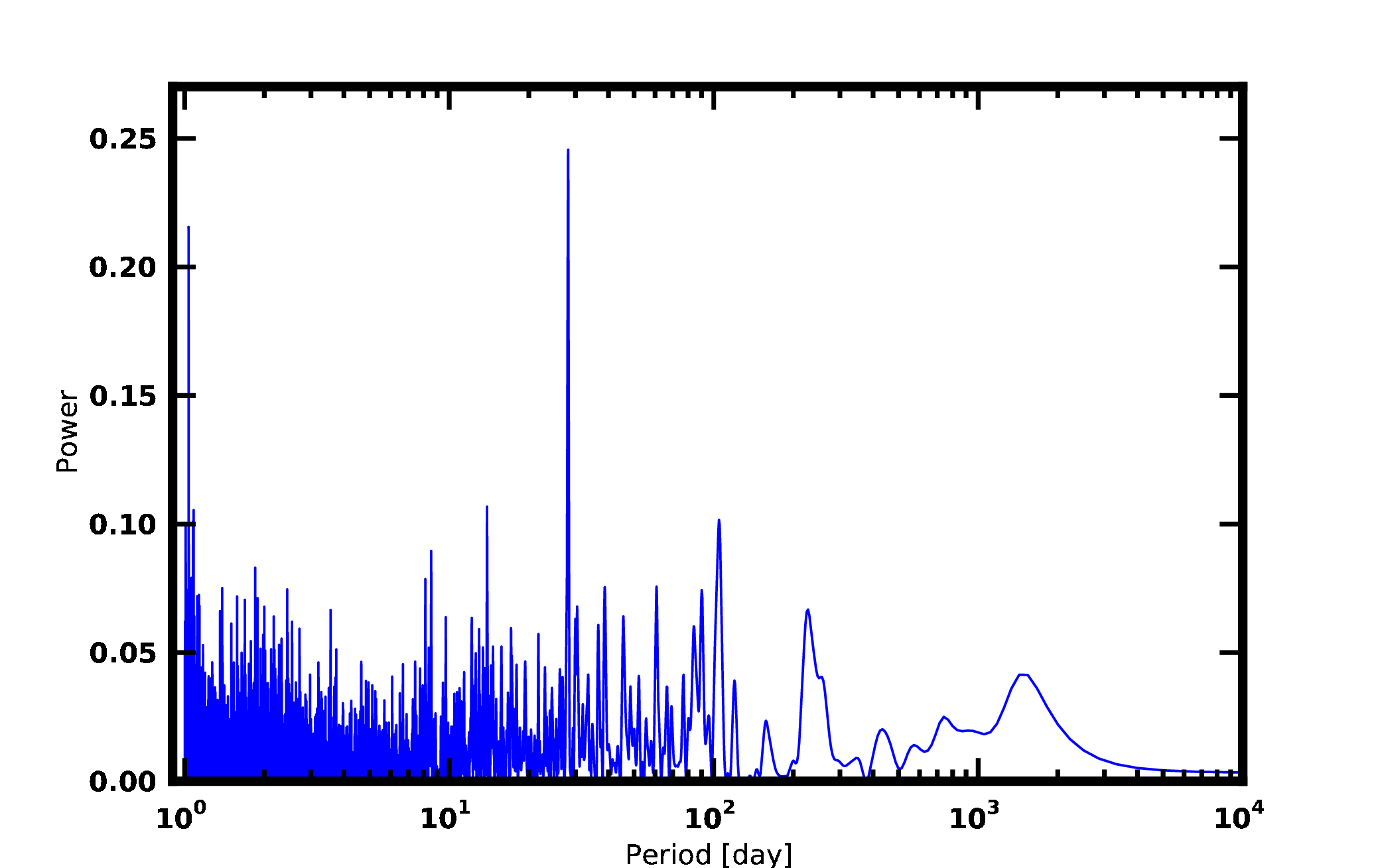}
\end{tabular}
\caption{Periodograms for the Gl~667C residual of radial velocity measurements
after subtraction of a drift plus a 2 planet solution. For all the
panels the period of the first planet is $P_b=7.2$~days. 
From the left to the right and from the top to
the bottom the period of the second planet is $P_c=28, 90, 106, 124, 186$
and $372$~days.}
\label{gl667_perio_2plan}
\end{figure*}

\subsection{A multi-keplerian plus a linear fit solution}

Following up on the periodogram of the residuals around the {\it 1
  planet + drift} model (Fig~\ref{gl667_perio_1plan} panel d), we observed
additional power excess around a series of periods, the 6 most
powerful peaks being around 28, 91, 105, 122, 185 and 364 d, with
$p=$0.225, 0.215, 0.185, 0.181, 0.134 and 0.123,
respectively. Bootstrap randomization indicates FAPs lower than
1/10,000 for the 4 most powerful peaks whereas Cumming's prescription
attributes them a FAP$< 10^{-12}$.

To model the RVs with {\it 2 planets + 1 drift}, one could pick the
most powerful peak as a guessed period and perform a local
minimization to derive all orbital parameters. We found it is however
not the most appropriate approach here because some of the less
powerful peaks actually correspond to a signal with high eccentricities,
and eventually turned to be better fit. To explore the best
solutions we generally prefer the global search implemented in $Yorbit$
with eccentricities left to vary freely. We identified several
solutions with similar $\sqrt{\chi2}$ values, reported in
Table~\ref{tab_sol_2plan}. For all solutions, the first planet and the
linear drift keep 
similar parameter values (see Sect~5.1). On the other hand, the different solutions
are discriminated by different orbital periods for the second planet,
with $P_c$ equals to $\sim$28, 90, 106, 124, 186 or 372 days.

\begin{figure*}
\begin{tabular}{cc}
(a) \includegraphics[width=8cm,angle=0]{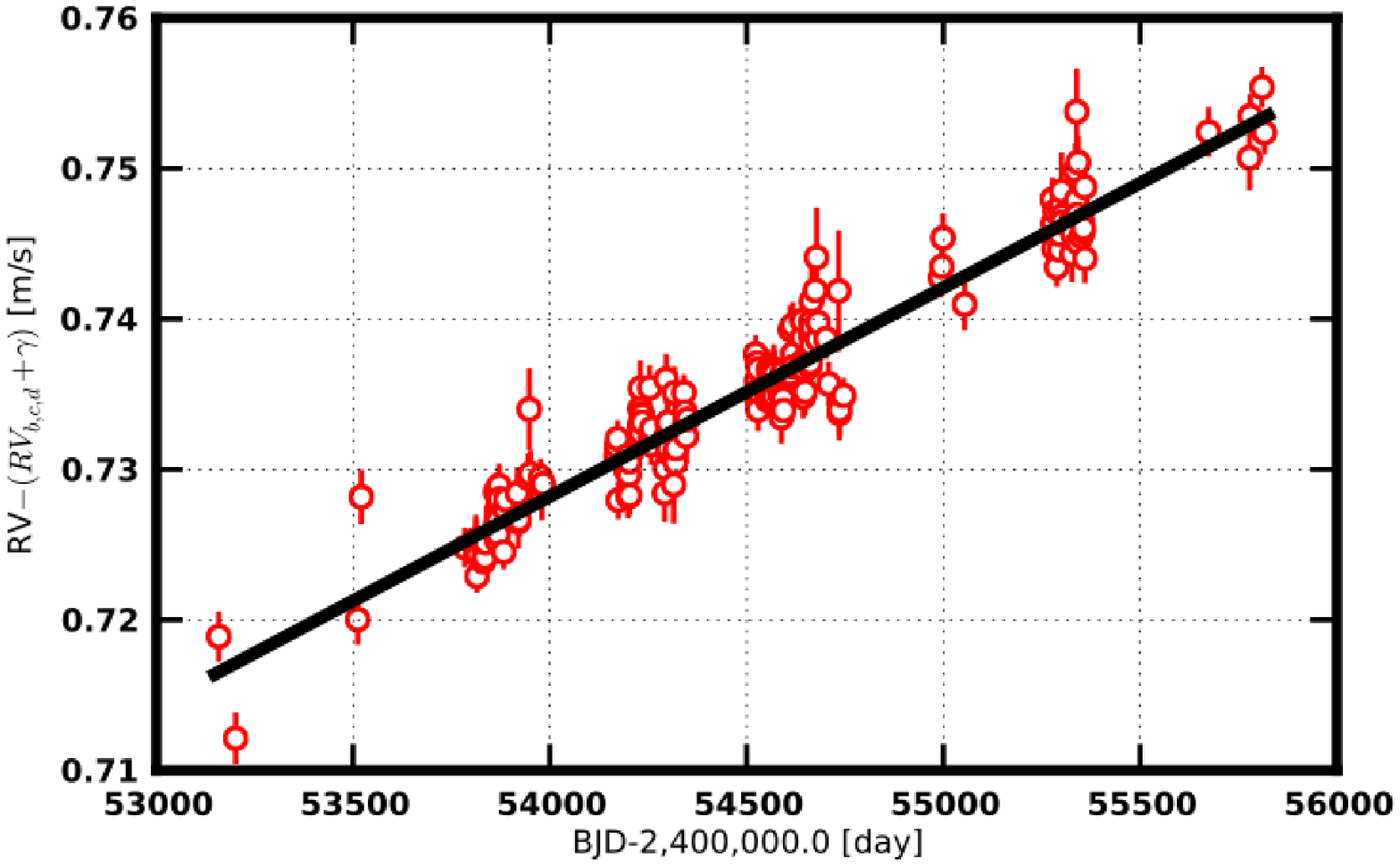} &
(b) \includegraphics[width=8cm,angle=0]{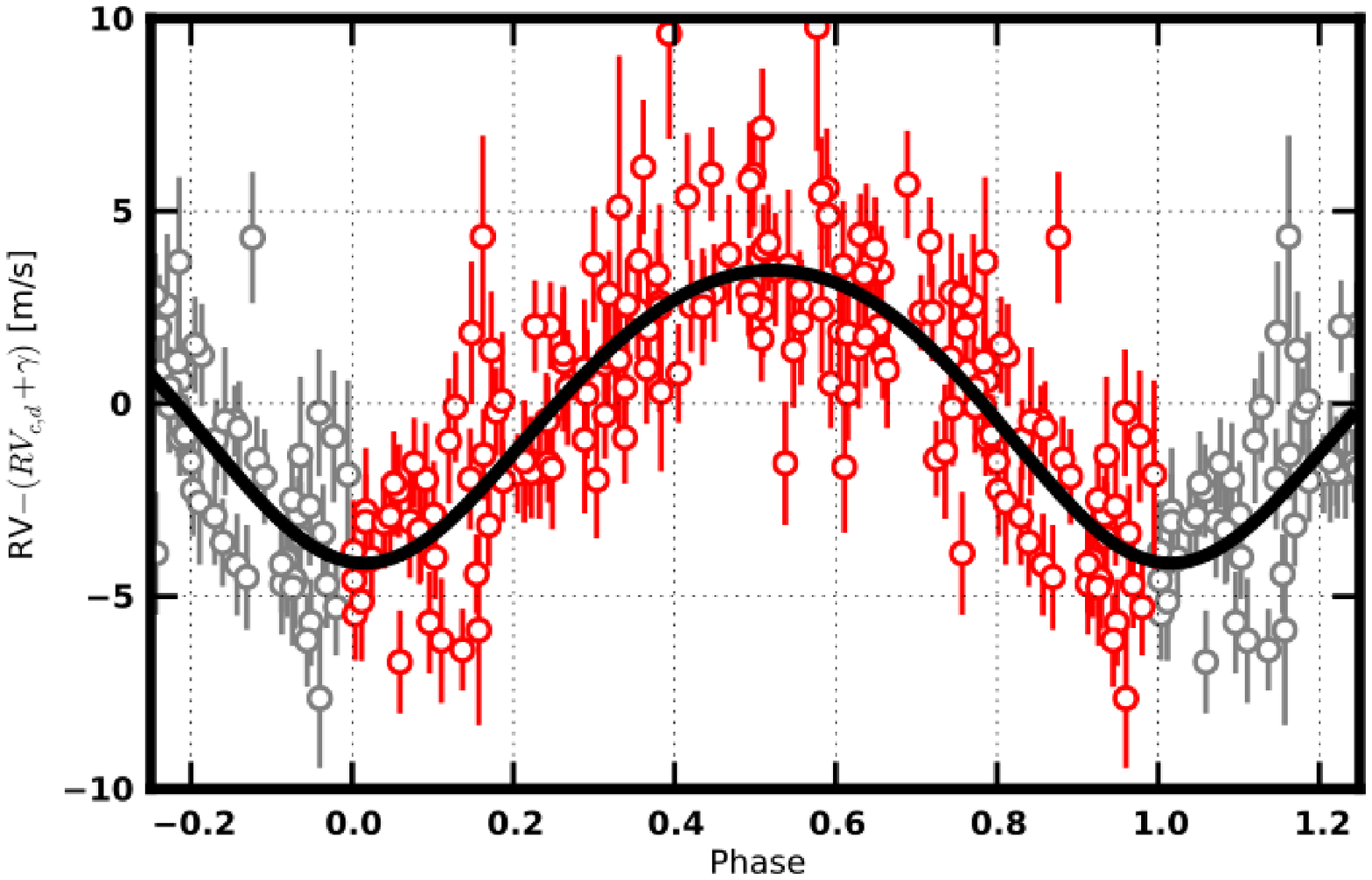} \\
(c) \includegraphics[width=8cm,angle=0]{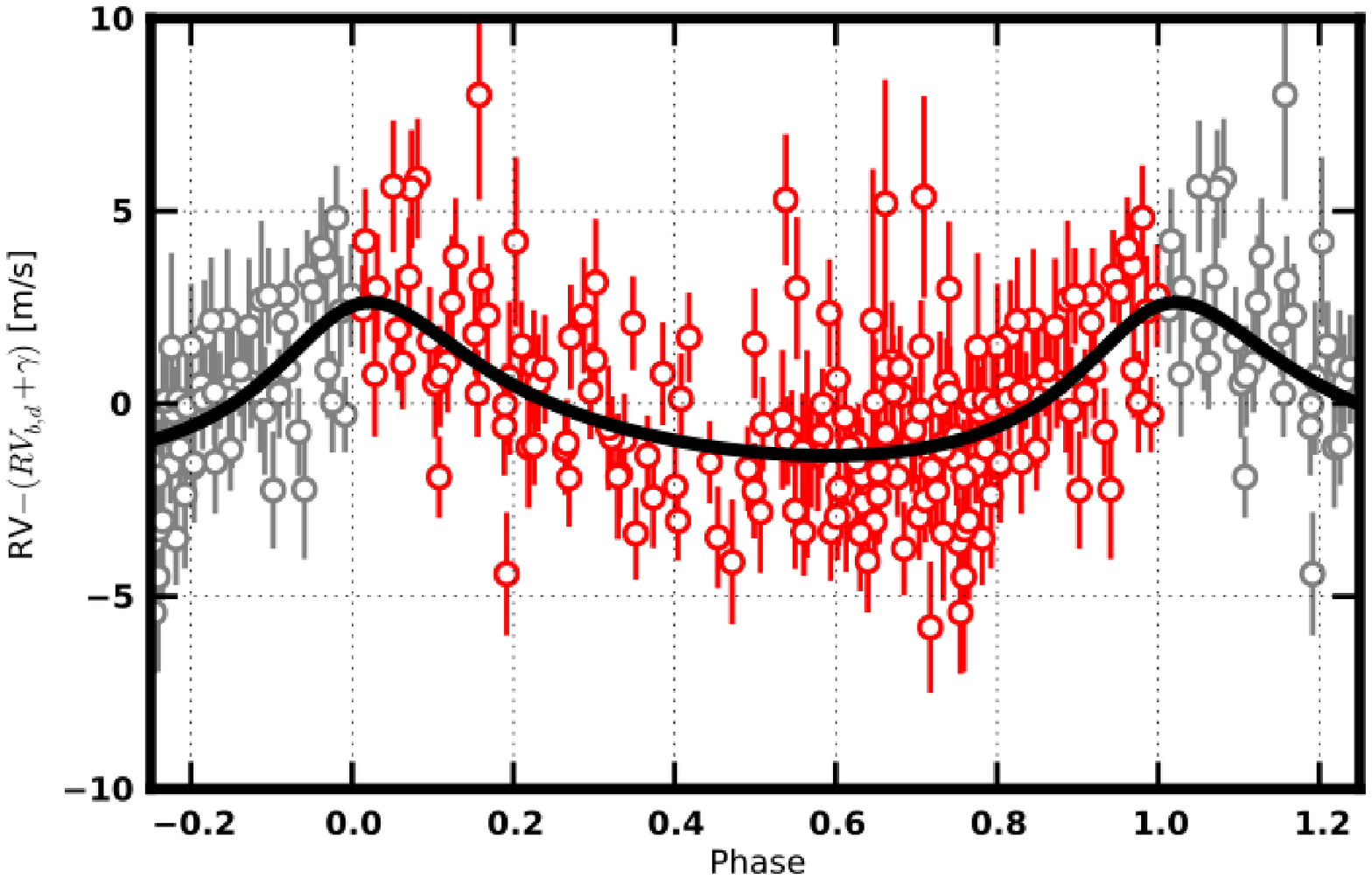} &
(d) \includegraphics[width=8cm,angle=0]{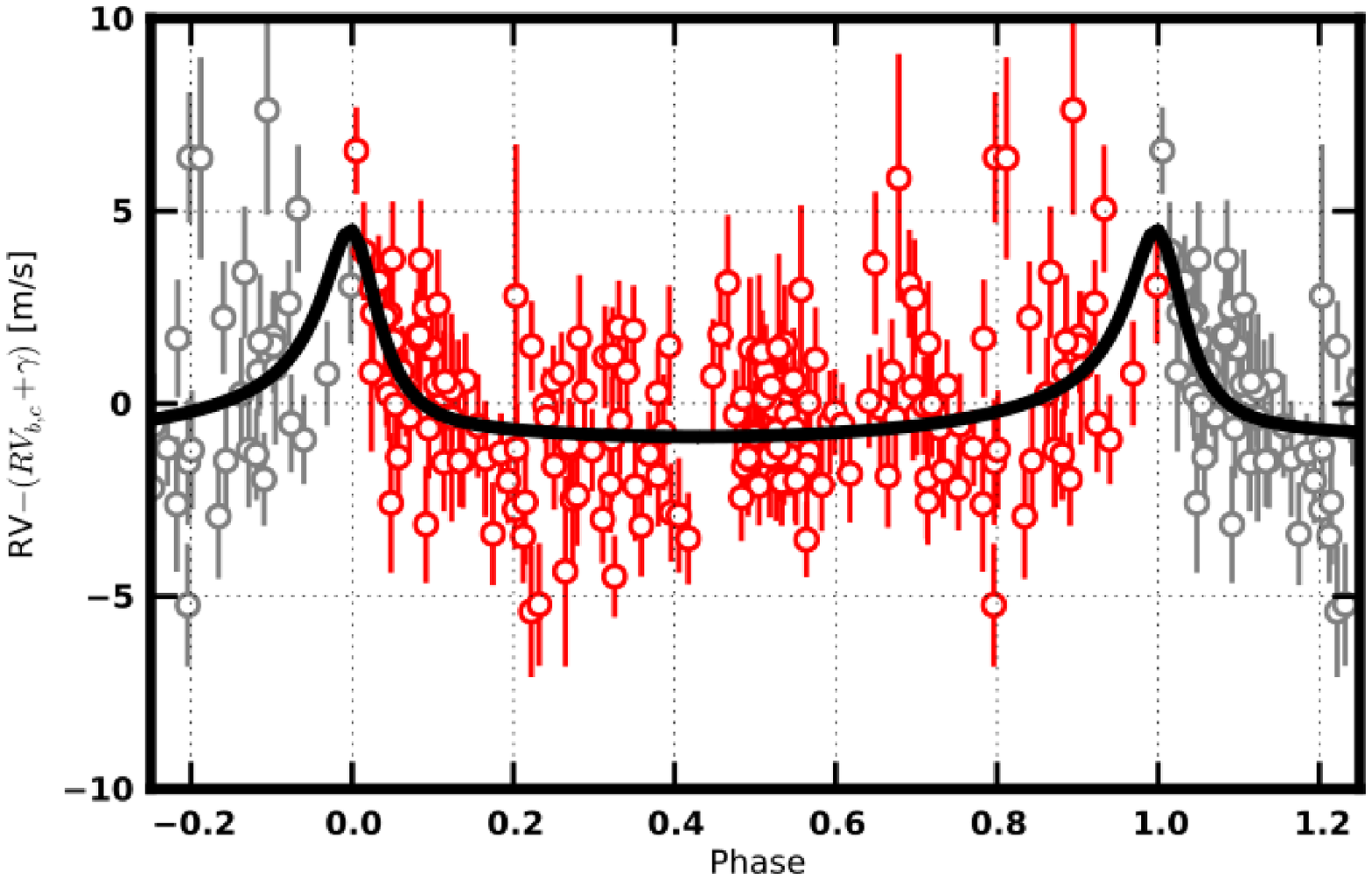} \\
(e) \includegraphics[width=8cm,angle=0]{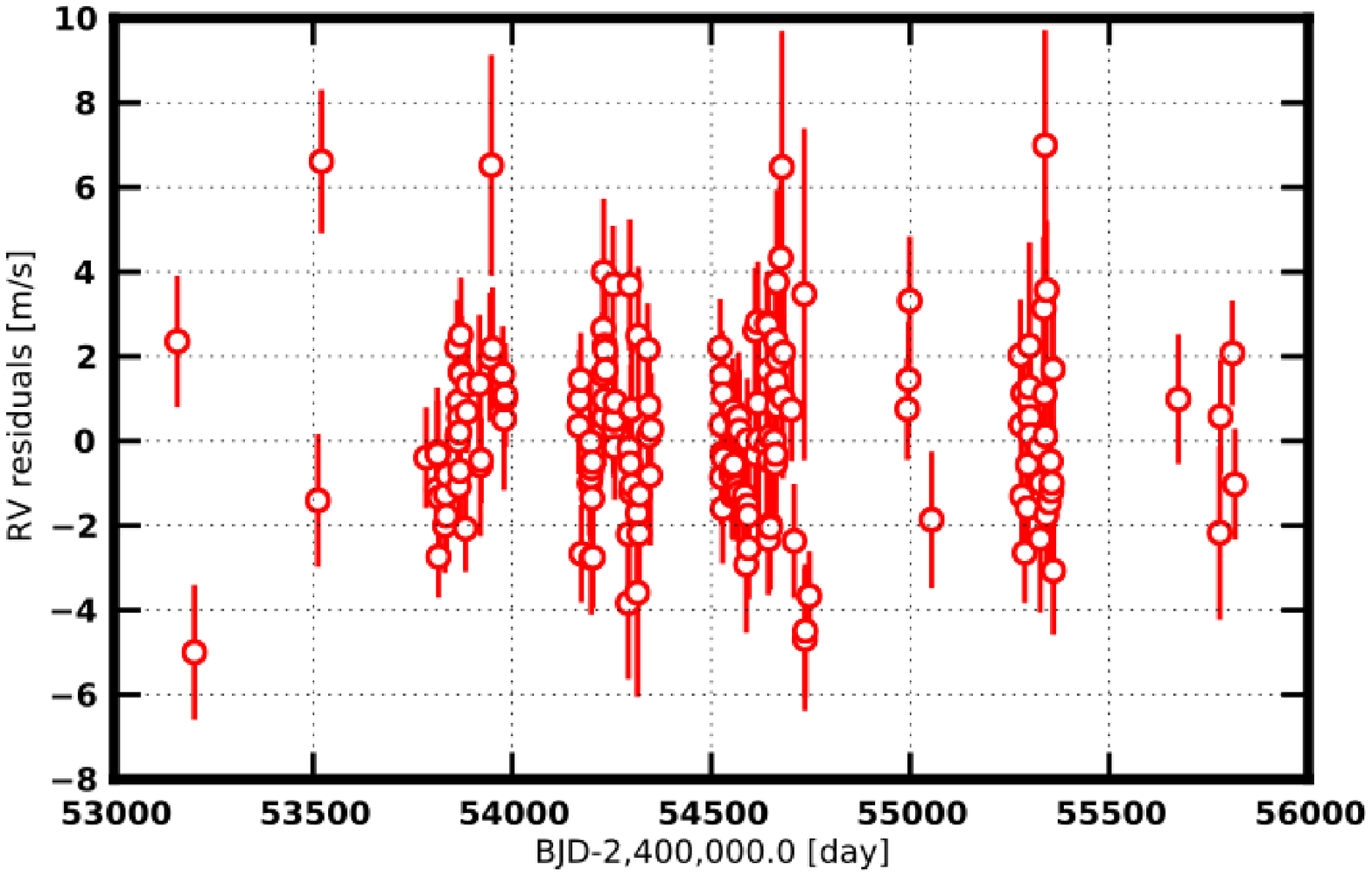} &
(f) \includegraphics[width=8cm,angle=0]{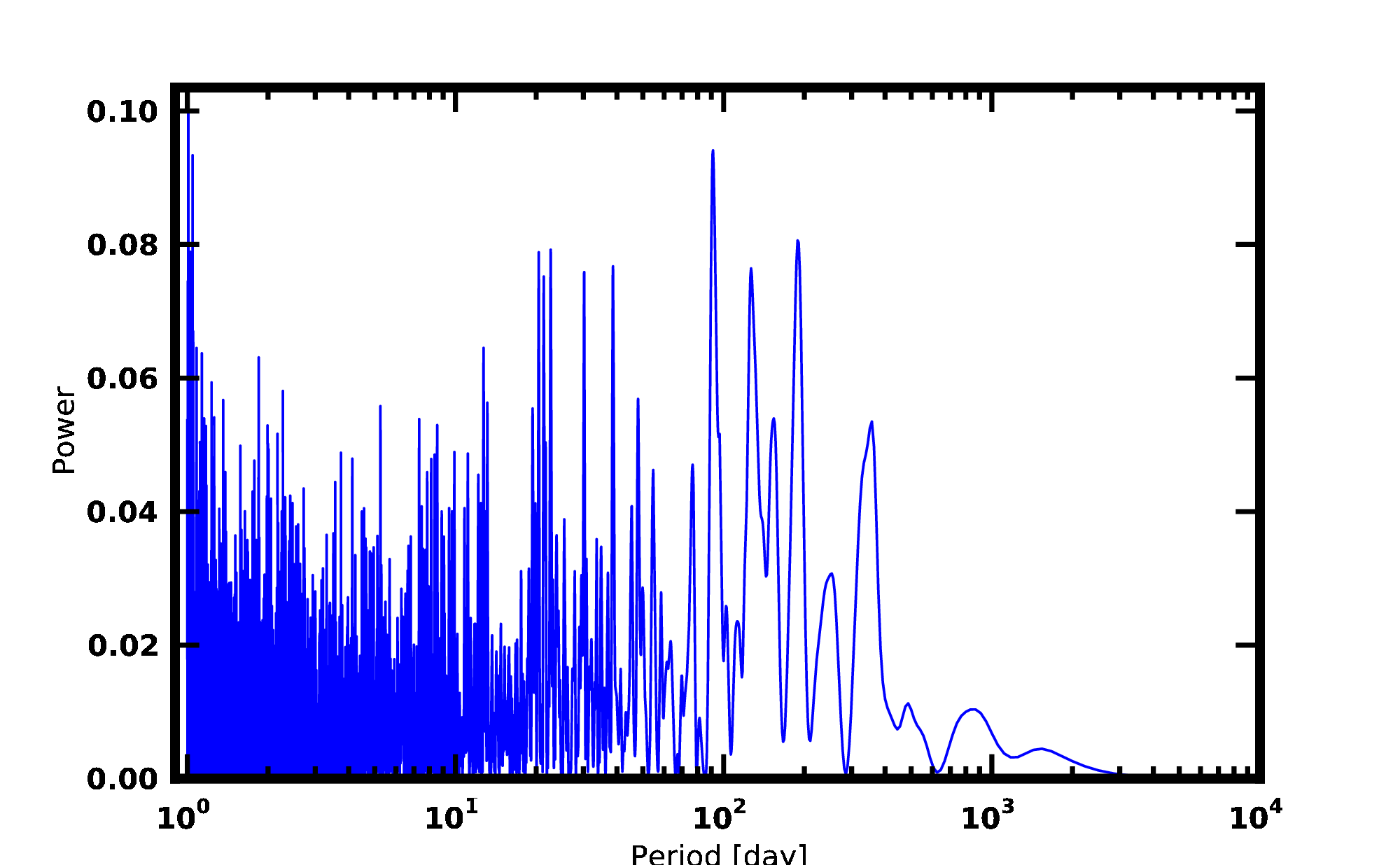}
\end{tabular}
\caption{Decomposition of our nominal 3 planet + 1 drift model for
  Gl~667C. (a) panel shows the contribution of the long-term drift
  with the 3 Keplerian contribution removed. Panels b, c and d show
  respectively radial velocity curves for keplerian $P_b=7.2$~d,
  $P_c=28.1$~d and $P_d=106.4$~d. Residual of this solution is showed
  in panel e with their periodogram in panel f.} 
\label{gl667_three_plan}
\end{figure*}

\begin{table*}
\caption{Six commensurable solutions (with similar reduced $\sqrt{\chi2}$) for the {\it 2-keplerian orbit $+$ 1 drift} model}
\begin{tabular}{cr@{$\pm$}lr@{$\pm$}lr@{$\pm$}lr@{$\pm$}lr@{}l}
\hline\hline 
Pl.	 &\multicolumn{2}{c}{$P$}  &\multicolumn{2}{c}{$K$}  &\multicolumn{2}{c}{$T_0$}	                 &\multicolumn{2}{c}{$e$}&\multicolumn{2}{c}{}\\
     &\multicolumn{2}{c}{[day]}&\multicolumn{2}{c}{[m/s]}&\multicolumn{2}{c}{BJD$-2,400,000.0$ [day]}&\multicolumn{2}{c}{$$} &\multicolumn{2}{c}{}   \\
\hline
  b     & 7.2002   &   0.0010 & 4.01 &     0.26 & 54443.33 &   0.53 & 0.13 &   0.06 &  & \\
  c     & 28.142   &   0.031   & 2.08 &     0.31 & 54461.3   &   1.6   & 0.32 &   0.12 &  &  \\
\multicolumn{3}{r}{$\gamma$}           &\multicolumn{2}{c}{=}& \multicolumn{2}{l}{$6.547 \pm 0.020 $}&\multicolumn{4}{l}{km/s}\\
\multicolumn{3}{r}{$\dot{\gamma} $} &\multicolumn{2}{c}{=}& \multicolumn{2}{l}{$1.80 \pm 0.13 $}&\multicolumn{4}{l}{m/s/yr}\\
\multicolumn{3}{r}{r.m.s.}                   &\multicolumn{2}{c}{=}& \multicolumn{2}{l}{$2.15$}                        &\multicolumn{4}{l}{m/s}\\
\multicolumn{3}{r}{$\sqrt{\chi2}$}                   &\multicolumn{2}{c}{=}& \multicolumn{2}{l}{$1.76\pm 0.05$}                        &\multicolumn{4}{l}{per degree of freedom}\\
\hline\\
  b     & 7.2002   &   0.0010 & 4.19 &     0.27 & 54442.89  &   0.38 & 0.17 &   0.06 &  & \\
  c     & 90.24     &   0.13     & 2.57&     0.61  & 54483.7    &   1.3   & 0.71 &   0.10 &  &  \\
\multicolumn{3}{r}{$\gamma$}           &\multicolumn{2}{c}{=}& \multicolumn{2}{l}{$6.548 \pm 0.019 $}&\multicolumn{4}{l}{km/s}\\
\multicolumn{3}{r}{$\dot{\gamma} $} &\multicolumn{2}{c}{=}& \multicolumn{2}{l}{$1.75 \pm 0.13 $}&\multicolumn{4}{l}{m/s/yr}\\
\multicolumn{3}{r}{r.m.s.}                   &\multicolumn{2}{c}{=}& \multicolumn{2}{l}{$2.17$}                        &\multicolumn{4}{l}{m/s}\\
\multicolumn{3}{r}{$\sqrt{\chi2}$}                   &\multicolumn{2}{c}{=}& \multicolumn{2}{l}{$1.77\pm 0.05$}                        &\multicolumn{4}{l}{per degree of freedom}\\
\hline\\
  b     & 7.2001   &   0.0010 & 3.90 &    0.25   &54442.93 &   0.53 & 0.13 &   0.06 &  & \\
  c     & 106.35   &  0.08      & 3.30 &     0.53  &54499.46 &   0.93 & 0.73 &   0.06 &  &  \\
\multicolumn{3}{r}{$\gamma$}           &\multicolumn{2}{c}{=}& \multicolumn{2}{l}{$6.548 \pm 0.019 $}&\multicolumn{4}{l}{km/s}\\
\multicolumn{3}{r}{$\dot{\gamma} $} &\multicolumn{2}{c}{=}& \multicolumn{2}{l}{$1.63 \pm 0.13 $}&\multicolumn{4}{l}{m/s/yr}\\
\multicolumn{3}{r}{r.m.s.}                   &\multicolumn{2}{c}{=}& \multicolumn{2}{l}{$2.07$}                        &\multicolumn{4}{l}{m/s}\\
\multicolumn{3}{r}{$\sqrt{\chi2}$}                   &\multicolumn{2}{c}{=}& \multicolumn{2}{l}{$1.69\pm 0.05$}                        &\multicolumn{4}{l}{per degree of freedom}\\
\hline\\
  b     & 7.2001   &   0.0010 & 3.99 &     0.27  & 54442.91 &   0.45 & 0.15  &   0.06   &  & \\
  c     &  123.98  &   0.14     & 3.05 &     0.64   &54481.38 &   0.88  &  0.80 &   0.06  &  &  \\
\multicolumn{3}{r}{$\gamma$}           &\multicolumn{2}{c}{=}& \multicolumn{2}{l}{$6.550 +- 0.019 $}&\multicolumn{4}{l}{km/s}\\
\multicolumn{3}{r}{$\dot{\gamma} $} &\multicolumn{2}{c}{=}& \multicolumn{2}{l}{$ 1.77 +- 0.13 $}&\multicolumn{4}{l}{m/s/yr}\\
\multicolumn{3}{r}{r.m.s.}                   &\multicolumn{2}{c}{=}& \multicolumn{2}{l}{$2.16$}                        &\multicolumn{4}{l}{m/s}\\
\multicolumn{3}{r}{$\sqrt{\chi2}$}                   &\multicolumn{2}{c}{=}& \multicolumn{2}{l}{$1.71\pm 0.05$}                        &\multicolumn{4}{l}{per degree of freedom}\\
\hline\\
  b     & 7.2001   &   0.0010 & 3.84 &    0.25    &54442.95 &   0.50 & 0.14  &   0.06  &  & \\
  c     &  186.08  &   0.30     & 2.79 &     0.47   &54604.4   &   1.3   &  0.80 &   0.05  &  &  \\
\multicolumn{3}{r}{$\gamma$}           &\multicolumn{2}{c}{=}& \multicolumn{2}{l}{$6.550 \pm 0.018 $}&\multicolumn{4}{l}{km/s}\\
\multicolumn{3}{r}{$\dot{\gamma} $} &\multicolumn{2}{c}{=}& \multicolumn{2}{l}{$1.75 \pm 0.12 $}&\multicolumn{4}{l}{m/s/yr}\\
\multicolumn{3}{r}{r.m.s.}                   &\multicolumn{2}{c}{=}& \multicolumn{2}{l}{$2.09$}                        &\multicolumn{4}{l}{m/s}\\
\multicolumn{3}{r}{$\sqrt{\chi2}$}                   &\multicolumn{2}{c}{=}& \multicolumn{2}{l}{$1.71\pm 0.05$}                        &\multicolumn{4}{l}{per degree of freedom}\\
\hline\\
  b     & 7.2001   &   0.0010 & 3.86 &     0.25  & 54442.96 &   0.49 & 0.14 &   0.06  &  & \\
  c     &  372.15  &   0.57     & 2.89 &     0.49   &54604.7   &   1.1   &  0.87 &   0.03 &  &  \\
\multicolumn{3}{r}{$\gamma$}           &\multicolumn{2}{c}{=}& \multicolumn{2}{l}{$6.550 \pm 0.018 $}&\multicolumn{4}{l}{km/s}\\
\multicolumn{3}{r}{$\dot{\gamma} $} &\multicolumn{2}{c}{=}& \multicolumn{2}{l}{$ 1.76 \pm 0.12 $}&\multicolumn{4}{l}{m/s/yr}\\
\multicolumn{3}{r}{r.m.s.}                   &\multicolumn{2}{c}{=}& \multicolumn{2}{l}{$2.09$}                        &\multicolumn{4}{l}{m/s}\\
\multicolumn{3}{r}{$\sqrt{\chi2}$}                   &\multicolumn{2}{c}{=}& \multicolumn{2}{l}{$1.71\pm 0.05$}                        &\multicolumn{4}{l}{per degree of freedom}\\
\hline
\end{tabular}
\label{tab_sol_2plan}
\end{table*}

To understand those signals we look at the periodograms of the
residuals around each solution. We found that, on the one hand, when
the solution is with $P_c=28$ d, the periodogram of the residual shows
the remaining peaks at 91, 105, 122, 185 and 364 days
(Fig.~\ref{gl667_perio_2plan} - panel a). On the other hand, when the
solution is with any of the other 
periods $P_c=90, 106, 124, 186$ or $372$ d then, only the 28 d peak
remains (e.g. Fig.~\ref{gl667_perio_2plan} - panels b, c, d, e and
f). This means that all peaks at 90, 
106, 120, 180 and 364 day actually correspond to a single signal, with
its harmonics and aliases, and that the peak around 28~d is another
independent signal.

One of the 4 signals identified so far (3 periodic signals + 1 linear
drift) have ambiguous solutions. Among the possible periods, the
$\sim$106-d period seems to correspond to the rotational period of the
star as seen in one activity indicator
(see. Sect. 5.3). Conservatively, we continue by assuming that signal
is due to activity and used the 106-d period to present our fiducial
solution. That assumption is given further credit when we run
$Yorbit$. Without any {\it a priori} on any signal, $Yorbit$ converges on a
solution with $P=7.2, 28$ and $106$~d. We present the orbital
parameters of a {\it 3 planet + 1 
  drift} model in Table~\ref{gl667_perio_3plan} and the RV
decomposition and  periodogram of
the residuals around that solution in Fig~\ref{gl667_three_plan}.   
That solution has rms and $\sqrt{\chi2}$ equal to 1.73 m/s and
1.44$\pm$0.06 per degree of freedom, respectively.

Finally, we inspect the residuals around that last model (see panels
(e) and (f) in Fig.~\ref{gl667_three_plan}). We found the
power maximum of the periodogram located around $P=1.0083$~d (a
possible 1-day alias with a period of 121 day) with power
$p_{max}=0.13$, to which we attributed a FAP of 2.8\%. Therefore, we did
not consider that significant sine signal remains in the
data. Nevertheless, remaining power excess around the period $\sim$90
and 122~d suggests the keplerian fit does not fully account for the
signal we attributed to stellar activity. 

\begin{table}
\caption{3 planets + 1 linear drift orbital solution}
\begin{tabular}{lllll} \hline
 & Gl~667Cb & Gl~667Cc & Gl~667Cd & \\ \hline 
$P$ [days]                  & 7.199$\pm$0.001 & 28.13$\pm$0.03 & 106.4$\pm$0.1 & \\
$e$                         & 0.09$\pm$0.05 & 0.34$\pm$0.10 & 0.68$\pm$0.06 & \\
$T0$ [JD - 2400000]         & 54443.1$\pm$0.6 & 54462$\pm$1 & 54499$\pm$1& \\
$\omega$ [deg]              & -4$\pm$33 & 166$\pm$20 & 7$\pm$8& \\
$K_1$ [m/s]                 & 3.8$\pm$0.2 & 2.0$\pm$0.3 & 2.7$\pm$0.4& \\
%$f(m)$ [$10^{13}$M$_{\odot}$] & 0.389 & 0.158 & 0.908 & \\ 
%$a_1.\sin{i}$ [$10^{5}$AU]   & 0.24730 & 0.45463 & 2.94581& \\
$M_2.\sin{i}$ [M$_{\oplus}$]  & 5.46 & 4.25 & 6.93 &  \\
$a$ [AU]                    & 0.0504 & 0.1251 & 0.3035 & \\ \hline
$\gamma$ [km/s]             & 6.55$\pm$0.02       &        &        & \\
$\dot{\gamma}$ [m/s/yr]     & 1.85$\pm$0.11 & & & \\
r.m.s. [m/s]                & 1.73 & & & \\
$\sqrt{\chi^2}$             & 1.44$\pm$0.06 & & & \\ \hline \hline
\end{tabular}
\label{gl667_perio_3plan}
\end{table}

\subsection{The planetary system around Gl~667C}

\begin{figure}
\includegraphics[width=9cm,angle=0]{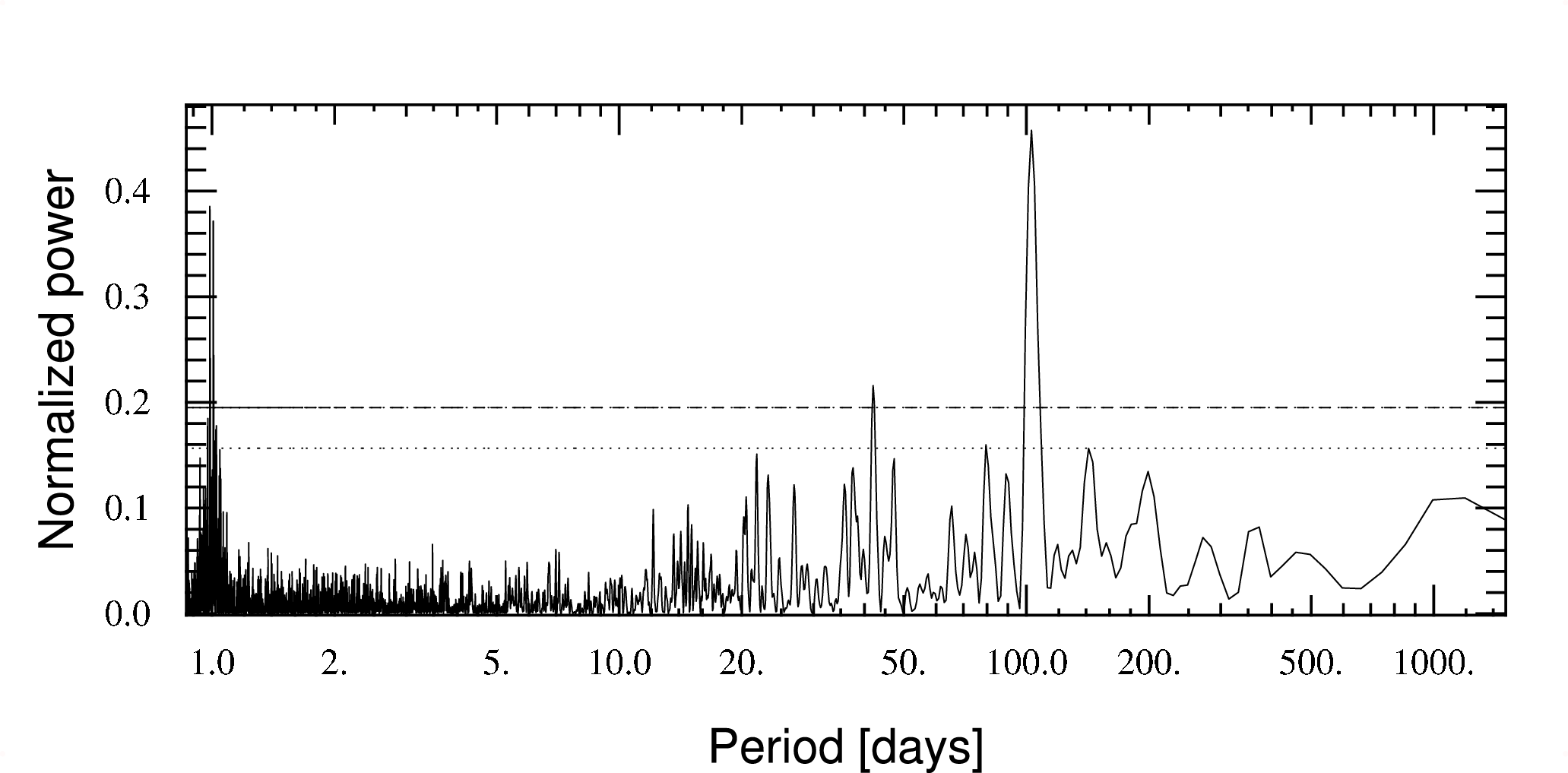}
\caption{Periodograms for the FWHM of the correlation peak for the Gl~667C 
HARPS measurements. A clear peak at $\sim$105 days is well visible and
is interpreted as the rotational period of Gl~667C. The line show the
10\% and 1\% FAP level.}
\label{per_fwhm}
\end{figure}

To assess if one of the radial velocity periodic signal could be due to
stellar rotation we searched for periodicity on several activity
diagnostics (H$\alpha$ and CaII-index, bisector-inverse slope (BIS) and
full-width at half-maximum (FWHM) of the cross-correlation function).
The clearest signal detected is a very high peak in the periodogram of the
FWHM at a period of $\sim$105 days with a FAP lower than 0.1\% (see
Fig.~\ref{per_fwhm}).

Periodic variation of the FWHM is an estimator of the rotation
\citep{queloz2009}. A rotational period of $\sim$105~days for Gl~667C
is consistent with the faint activity level of these stars, discussed
in Sect.~3.2 and such period matches one of the power excess seen in our
radial velocity periodogram analysis. Thus our favorite explanation is
that the period of
$\sim$105~days and all the harmonics or alias at $P=90, 124, 186$
and $372$~days (see Sect. 5.2) are due to stellar rotation. However, 
we tried to use different sub-sets of RV points and found that the
best solution may appear with $P_d=90, 106$ or $186$~d.
We therefore consider that the RV signal at large ($P>90$)
period is not definitively assigned to date, its origin may be stellar
rotation (the most probable in our point of view), or a high eccentricity
planet, 2 planets in resonnance or a combination of several of these
explanations. 

The RV signal at 7.2 and 28.1~d are completely
independent of the large period ones and cannot be due to stellar
rotation. Thus at least 2 super-Earths
are present at close separation of Gl~667C with a mass of 5.5 and
4.25~M$_{\oplus}$. Already announced by our team in
\citet{bonfils2011} with GTO-HARPS data, the use of supplementary
HARPS RV allows us to specify their parameters. At a separation
of 0.05 and  0.12 a.u. from their star, Gl~667Cb and Gl~667Cc are
respectively illuminated by a bolometric flux, per surface unity, 5.51
and 0.89 times than what the Earth receives from the Sun (in using
stellar luminosity of Table~\ref{tab_parametre}). The super-Earth
Gl~667Cc is then in the middle of the habitable zone of its star. We
discuss this point in detail in the next section.

\section{A planet in the middle of the habitable zone of an M dwarf}

\subsection{Gl~667Cc}

Super-Earths in the habitable zone of their host stars, 
which by definition is the region where liquid water can 
be stable on the surface of a rocky planet 
\citep{huang1959, kasting93}, currently garner considerable
interest. For a detailed discussion of the HZ we 
refer the reader to \citet{selsis2007} or \citet{kaltenegger2011a},
but we summarize the most salient points.
One important consideration is that a planets with masses
outside the 0.5--10$M_{\oplus}$ range cannot host liquid surface water.
Planets under the lower end of this range have too weak a gravity 
to retain a sufficiently dense atmosphere, and those above the upper
end accrete a massive He-H envelope. In either case, the pressure 
at the surface is incompatible with liquid water. The 10$M_{\oplus}$ 
upper limit is somewhat fuzzy, since planets in the 3--10$M_{\oplus}$ range
can have very different densities (reflecting different structure) for
a given mass: Earth-like, Neptune-like, and Ocean planets can all
exist for the same mass \citep[e.g. Fig. 3 in][]{winn2011}. 

To potentially harbor liquid water, a planet with a dense atmosphere 
like the Earth needs an equilibrium temperature between 175K and 270K 
\citep[][and references therein]{selsis2007, kaltenegger2011a}. 
If $T_{eq} > 270$K, a planet with a water-rich atmosphere will 
experience a phase of runaway greenhouse effect
\citep[see][]{selsis2007}. At the
other end of the range, CO$_2$ will irreversibly freeze out from 
the atmosphere of planets with $T_{eq} < 175$K, preventing a sufficient 
greenhouse effect to avoid freezing of all surface water. Either case 
obviously makes the planet uninhabitable. The inner limit of the 
habitable zone is well constrained. The outer one, by contrast, is 
very sensitive to the complex and poorly constrained meteorology
of CO$2$ clouds, through the balance between their reflecting
efficiency (which cools the planet) and their greenhouse effect 
(which warms it) \citep{forget1997,mischna2000,selsis2007}. 
Finally, a location inside the habitable zone is a necessary
condition for hosting surface water, but by no means a sufficient
one: long term survival of surface water involves complex ingredients 
such as a carbonate-silicate geological cycle, and an adequate
initial H$_2$O supply.

Adopting the notations of \citet{kaltenegger2011b} the equilibrium
temperature of a planet is:

\begin{equation}
T_{eq} = ((1-A)L_{star}/(4{\beta}D^2))^{1/4}
\end{equation}

where $\beta$ is the geometrical fraction of the planet surface 
which re-radiates the absorbed flux ($\beta = 1$ for a rapidly 
rotating planet with a dense atmosphere, like the Earth; 
$\beta = 0.5$ for an atmosphere-less planet that always 
presents the same side to its star) and $A$
the wavelength-integrated Bond albedo. 
Using the stellar parameters of Table~{\ref{tab_parametre}}, the
equilibrium temperature of Gl433b and Gl667Cb are respectively
495$(\frac{1-A}{\beta})^{1/4}$K and
426$(\frac{1-A}{\beta})^{1/4}$K. Both planets are firmly in the 
hot super-Earth category. Gl~667Cc, by contrast, has 
$T_{eq}~=~270(\frac{1-A}{\beta})^{1/4}$K, to be compared to
$T_{eq}~=~278(\frac{1-A}{\beta})^{1/4}$K for the Earth. 

Recent 3-D atmospheric models \citep{heng2011b, heng2011a} 
suggest that Gl667Cc most likely has $\beta$ close to 1,
even in the case it is tidally locked and always shows the 
same hemisphere to its star, since they find that a modestly 
dense Earth-like atmosphere ensures a full re-distribution 
of the incoming energy. That conclusion, {\it a fortiori}, also
applies for the presumably denser atmosphere of a 4.25M$_{\oplus}$ 
planet \citep{wordsworth2011}. With $T_{eq} \sim 270(1-A)^{1/4}$K, 
Gl667Cc is therefore in the HZ for any albedo in the [0-0.83]
range.

Detailed 3-dimensional atmospheric simulations 
\citep[e.g.][]{wordsworth2011,heng2011c} will need to be
tuned to the characteristics of Gl~667Cc to ascertain its
possible climate. In the mean time, this planet, which receives from 
its star 89\% of the bolometric solar flux at Earth, is a very
strong habitable planet candidate.

\subsection{Particularity of the habitable zone around M dwarfs}

Two main differences between planets in HZ
of solar like stars and M dwarfs are often pointed out. 
Firstly, a planet in HZ of an M dwarf is
closer to its star, and therefore subject to more intense 
tidal forces. As a result, it is likely to quickly be captured 
into a spin-orbit resonance. The Solar System demonstrates, 
however, that this does not necessarily imply that it will
be forced into synchronous rotation. The final 
equilibrium rotation of a tidally influenced planet 
depends on both its orbital eccentricity and the density of its 
atmosphere \citep{doyle1993, correia2008, correia2010, heller2011}. 
Mercury, for instance, has been captured into the 3:2,
rather than 1:1, spin-orbit resonance \citep{correia2004}, and 
Venus has altogether escaped capture into a resonance because 
thermal atmospheric tides counteract its interior tides 
\citep{correia2003}. Whatever the final spin-orbit ratio, the 
tidal forces will influence the night and day succession, and 
therefore the climate. As discussed above however, energy
redistribution by an atmosphere at least as dense as that 
of the Earth is efficient \citep{wordsworth2011, heng2011a,
heng2011b}, and will prevent glaciation and atmospheric 
collapse on the night side.

The second major difference is stellar magnetic activity, through its
dependence on stellar mass. M dwarfs, on average, are much more active 
than solar-like star. This results from the compounding of two effects: 
lower mass stars have much longer braking times for stellar rotation 
\citep{delfosse98, barnes03, delorme11}, and for the same rotation
period they are more active \citep{kiraga2007}. As a result, 
a planet in HZ of an early M dwarf receives
intense X and UV radiation for 10 times longer than the 
$\sim$100~Myrs which the solar system spent close to a very
active Sun \citep{ribas2005, selsis2007}. Stellar X and UV 
radiation, as well as coronal mass ejections (CMEs), could
potentially cause a major planetary atmosphere escape 
\citep[see][for reviews]{scalo2007, lammer2009}.  

Several theoretical studies have looked into the evolution of
planetary atmospheres in M-dwarf HZ: do they lose essential 
chemical species, like H$_2$O, and can they even be completely 
eroded out? The problem is complex, with many uncertain parameters,
and the span of possible the answers is very wide. Amongst other 
uncertain factors, the escape ratio sensitively depends on: (1) the intensity 
and frequency of CMEs for M dwarfs (which is poorly known, and may be 
much smaller that initialy claimed \citep{vandenoord1997,wood05}),
(2) the magnetic moment of the planet (a strong magnetosphere 
will protect the planet against CME-induced Ion Pick
Up \citep{lammer2007}), and (3) details of the atmospheric chemistry  
and composition (through atmosphere-flare photo-chemical interaction 
\citep{segura10} and IR radiative cooling from vibrational-rotational band
\citep{lammer2007}). In two limiting cases of planets in M-dwarfs HZ, 
\citet{lammer2007} conclude that the atmosphere of an unmagnetized 
Earth-mass planet can be completly eroded during the Gyr-long active 
phase of its host star, while \citet{tian09} find that the atmosphere
of a 7 Earth-mass super-Earth is stable even around very active
M dwarfs, especially if that atmosphere mostly contains CO$_2$. 
Continuous outgassing of essential atmospheric species
from volcanic activity can, of course, also protect an atmosphere 
by compensating its escape.

These differences imply that a planet in the habitable zone of
an M dwarf is unlikely to be a twin of the Earth. Habitability 
however is not restricted to Earth twins, and \citet{barnes10} 
conclude that ``no known phenomenon completely precludes the 
habitability of terrestrial planets orbiting cool stars.''
A massive telluric planet, like Gl667Cc ($M_2.\sin{i}~=~4.25~M_{\oplus}$), 
most likely has a massive planetary core, and as a consequence a
stronger dynamo and a more active volcanism. Both factors help
protect against atmospheric escape, and super-Earths may 
perhaps be better candidates for habitability around M dwarfs 
than true Earth-mass planets.

\subsection{Gl~667Cc compared with the another known planets in the
  habitable zone} 

HARPS has previously discovered two planets inside HZ.
Gl~581d \citep{udry07,mayor09} ($M_2.\sin{i} = 7~M_{\oplus}$) 
receives from its M3V host star just $\sim$~25\% of the 
energy that the Earth receives from the Sun, and is
thus located in the outer habitable zone of its star. 
Recent detailed models \citep{wordsworth2011} confirm 
that Gl~581d can have surface liquid water for a wide range of
plausible atmospheres. HD85512b \citep{pepe2011} is a 
$M_2.\sin{i}~=~3.5~M_{\oplus}$ planet in the inner habitable
zone of a K5-dwarf, and receives $\sim$twice as much stellar 
energy as the Earth. It can harbor surface liquid water if it
is covered by at least 50\% of highly reflective clouds 
\citep{kaltenegger2011b}.

\citet{vogt2010} announced another super-Earth in 
the HZ of Gl~581, which they found in an analysis combining 
HARPS and HIRES radial velocity data. The statistical significance 
of that detection was, however, immediately questioned 
\citep[e.g.][]{tuomi2011}, and an extended HARPS dataset now 
demonstrates that the planet is unlikely
to exist with the proposed parameters \citep{forveille2011}.

\citet{borucki2011a} announced 6 planetary candidates in the HZ of
{\it Kepler} targets with a radius below twice that of Earth,
which they adopt as a nominal limit between telluric and Neptune-like
planets. \citet{kaltenegger2011a} reduce this number to 3 HZ planetary
candidates, by discarding 3~planets which they find are too hot
to host water if their coverage by reflective clouds is under 50\%.
\citet{borucki2011b} recently confirmed Kepler~22b (listed as 
KOI-87.01 in \citet{borucki2011a}) as a planet, making it
the first planet with a measured radius orbiting in a habitable zone. 
\citet{kaltenegger2011a} however discarded Kepler~22b from their
candidate list because its radius is above 2 Earth-radii.

To summarize, two planets with measured minimum masses in the
range for telluric planets (Gl~581d and HD85512b) are know to 
orbit in habitable zones. Another one, with a measured radius 
slightly above the nominal limit for a rocky planet, orbits 
in a similar location (Kepler~22b). Finally, 3 {\it KEPLER} 
candidates with radii corresponding to telluric planets
and positions in habitable zones currently await confirmation. 
Gl~667Cc, announced in \citet{bonfils2011} and discussed in 
detail here, is actually the most promising of those for holding
conditions compatible with surface liquid water. Receiving
$\sim$10\% less stellar energy than the Earth and with a minimum mass
of $M_2.\sin{i} = 4.25~M_{\oplus}$, it is very likely to be a
rocky planet in the middle of the habitable zone of its star.

\section{Summary and conclusions}

In this paper we analysed in detail 3 super-Earths announced in
\citet{bonfils2011} : Gl~433b, Gl~667Cb and Gl~667Cc. One, Gl667Cc, 
is a $M_2.\sin{i} = 4.25~M_{\oplus}$ planet in the middle of the
habitable zone of a M1.5V star. It is to date the known 
extra-solar planet with characteristics closest to Earth,
but does not approach being an Earth twin. The main differences 
from Earth are a significantly higher mass and a different stellar
environment, which potentially can have caused divergent evolutions.

Host stars of giant planets are preferentially metal-rich
\citep[e.g.][]{santos2001, santos2004, fischer2005}. The detection
rate of planets with masses under 30-40M$_{\oplus}$, by contrast, 
does not clearly correlate with stellar metallicity \citep{mayor2011}. 
This context makes the discovery of two super-Earths around Gl~667C 
([Fe/H]$\sim$-0.6, Sect.~3) less surprising, but it remains one of 
the most metal-poor planetary host known to this date, with just 
6 planetary host stars listed with a lower metallicity in the Extrasolar 
Planets Encyclopaedia\footnote{http://http://exoplanet.eu/index.php}
\citep{schneider2011}. The metallicity of Gl~433, [Fe/H]$\sim$ -0.2,
is close to the median value of solar neighbourhood, and therefore
unremarkable.

Gl~667Cb and c orbit the outer component of a hierarchical triple 
system. Fewer than 10 other planetary systems, listed in 
\citet{desidera2011}, share that characteristics. These authors
conclude that planets occur with similar frequencies around the isolated
component of a triple system and around single stars. The planets around 
Gl~667C, however, are unusual in orbiting around the lowest-mass 
component of the system.

The three planets discussed here enter in our \citet{bonfils2011} 
statistical study, which establishes that super-Earths are very 
common around M dwarfs. Our HARPS survey of $\sim$100 stars has,
in particular, found two super-Earths in habitables zones, Gl~581d 
and Gl~667Cc, even though both planets are located in parts of 
the mass-period diagram where its detection completeness is 
under 10 percent. This clearly indicates that super-Earths
are common in the habitable zones of M dwarfs, with \citet{bonfils2011} 
a 42$^{+54}_{-13}\%$ frequency. Future instruments optimized
for planet searches around M dwarfs, like the SPIRou (on CFHT) 
and CARMENES (at Calar Alto observatory) near-IR spectrographs, 
will vastly increase our inventory of such planets around
nearby M dwarfs. Such instrument should be able to identify 
$\sim$50-100 planets in the habitable zones of M-dwarfs, 
and with a 2-3\% transit probability for those, to find at
least one transiting habitable planet around a bright M dwarf.
Such {\it radial-velocity educated approach} is already undertaken
with HARPS \citep{bonfils2011b}. In this context
we stress that a photometric search for transits must be carried out
shortly for Gl~667Cc, which have a 2.5\% probability to occur.

\begin{acknowledgements}
We thank the 3.6-m team for their support during the observations 
which produced these results. We thank Abel M\'endez for constructive
discussion about the ``habitability'' of Gl~667Cc. Financial support
from the "Programme National de Plan\'etologie'' (PNP)                
of CNRS/INSU, France, is gratefully acknowledged. NCS acknowledges the
support from the European Research Council/European Community under
the FP7 through Starting Grant agreement number 239953, and from
Funda\c{c}\~ao para a Ci\^encia e a Tecnologia (FCT) through program
Ci\^encia\,2007, funded by FCT (Portugal) and POPH/FSE (EC), and in
the form of grants reference PTDC/CTE-AST/098528/2008 and
PTDC/CTE-AST/098604/2008. V.N. would also like to acknowledge the
support from the FCT in the form of the fellowship SFRH/BD/60688/2009. 

\end{acknowledgements}

\bibliographystyle{aa}

\bibliography{biblio_plan}

\appendix

\section{Radial velocity measurements}

\begin{table}
\caption{Radial-velocity measurements and errors bars for Gl~433. All
  values are relative to the solar system barycenter, the secular
  acceleration is not substracted (see Sect.~2).}
\begin{tabular}{lll} \hline \hline
JD-2400000 & RV & Uncertainty \\ 
           & [km.s$^{-1}$] & [km.s$^{-1}$] \\ \hline
52989.835057	& 18.167720	& 0.001660 \\
52996.843978	& 18.164793	& 0.001320 \\
53511.574791	& 18.161194	& 0.001000 \\
53516.592190	& 18.160386	& 0.001400 \\
53516.597028	& 18.160016	& 0.001680 \\
53520.601574	& 18.170608	& 0.002200 \\
53809.751435	& 18.164037	& 0.000900 \\
53810.729535	& 18.160597	& 0.000810 \\
53817.770817	& 18.162430	& 0.000960 \\
54134.827289	& 18.162990	& 0.000870 \\
54200.675734	& 18.165507	& 0.001000 \\
54229.642760	& 18.165469	& 0.001470 \\
54256.554461	& 18.170100	& 0.000960 \\
54257.531401	& 18.171791	& 0.001100 \\
54258.495969	& 18.168751	& 0.001000 \\
54296.563423	& 18.167497	& 0.001410 \\
54299.536021	& 18.167428	& 0.001700 \\
54459.849816	& 18.161704	& 0.000910 \\
54460.851730	& 18.159534	& 0.001190 \\
54526.745535	& 18.161241	& 0.000860 \\
54549.673279	& 18.168181	& 0.001440 \\
54552.694028	& 18.168932	& 0.000950 \\
54556.639741	& 18.166923	& 0.001140 \\
54562.685147	& 18.163146	& 0.000830 \\
54566.649449	& 18.169138	& 0.000840 \\
54570.611066	& 18.161279	& 0.001050 \\
54639.547895	& 18.165018	& 0.001150 \\
54640.536378	& 18.169428	& 0.001010 \\
54641.508220	& 18.168608	& 0.000880 \\
54642.531642	& 18.166339	& 0.001010 \\
54643.541084	& 18.164339	& 0.001120 \\
54645.510158	& 18.164340	& 0.001220 \\
54646.510318	& 18.169850	& 0.001070 \\
54647.470551	& 18.171231	& 0.001030 \\
54648.507247	& 18.171541	& 0.000960 \\
54658.462295	& 18.164905	& 0.001010 \\
54660.460084	& 18.166296	& 0.001280 \\
54661.462908	& 18.166917	& 0.001080 \\
54662.473787	& 18.168927	& 0.001070 \\
54663.461670	& 18.167557	& 0.000860 \\
54664.466903	& 18.166128	& 0.001580 \\
54665.470990	& 18.163598	& 0.001000 \\
54666.465274	& 18.166129	& 0.000900 \\
54672.514921	& 18.163801	& 0.001190 \\
54674.470816	& 18.160382	& 0.001300 \\
54677.477550	& 18.166513	& 0.001150 \\
54678.473044	& 18.171103	& 0.001990 \\
54679.476248	& 18.165294	& 0.001050 \\
54681.469000	& 18.166555	& 0.001200 \\
54682.474358	& 18.167625	& 0.000940 \\
55041.491690	& 18.165273	& 0.000970 \\
55046.463763	& 18.168735	& 0.001090 \\
55047.482719	& 18.169075	& 0.001310 \\
55048.480602	& 18.164995	& 0.001370 \\
55049.486609	& 18.162866	& 0.001340 \\
55050.478302	& 18.162216	& 0.002460 \\
55053.476079	& 18.163187	& 0.001160 \\
55054.481130	& 18.165818	& 0.001710 \\
55055.465316	& 18.164458	& 0.001300 \\
55056.458633	& 18.162689	& 0.001150 \\
55057.474960	& 18.161779	& 0.001320 \\
55234.724210	& 18.162092	& 0.001120 \\
55887.861310	& 18.167640	& 0.001230 \\
55889.846375	& 18.160361	& 0.001660 \\
55891.844384	& 18.164192	& 0.001080 \\
55892.857129	& 18.165482	& 0.001120 \\
55893.838113	& 18.164223	& 0.001060 \\ \hline
\end{tabular}
\end{table}

\begin{table}
\caption{Radial-velocity measurements and errors bars for Gl~667C
  (first part). All
  values are relative to the solar system barycenter, the secular
  acceleration is not substracted (see Sect.~2).}
\begin{tabular}{lll} \hline \hline
JD-2400000 & RV & Uncertainty \\ 
           & [km.s$^{-1}$] & [km.s$^{-1}$] \\ \hline
53158.764366	& 6.543610	& 0.001490 \\
53201.586793	& 6.535135	& 0.001530 \\
53511.798846	& 6.538483	& 0.001510 \\
53520.781048	& 6.546518	& 0.001630 \\
53783.863348	& 6.547189	& 0.001130 \\
53810.852282	& 6.543955	& 0.001000 \\
53811.891816	& 6.547456	& 0.001470 \\
53812.865858	& 6.546566	& 0.001200 \\
53814.849083	& 6.537267	& 0.000910 \\
53816.857459	& 6.537278	& 0.001130 \\
53830.860468	& 6.539786	& 0.001060 \\
53832.903068	& 6.546288	& 0.001150 \\
53834.884977	& 6.545569	& 0.001040 \\
53836.887788	& 6.541110	& 0.000910 \\
53861.796371	& 6.554704	& 0.001060 \\
53862.772051	& 6.554435	& 0.001210 \\
53863.797178	& 6.552085	& 0.001020 \\
53864.753954	& 6.549526	& 0.001120 \\
53865.785606	& 6.544717	& 0.001010 \\
53866.743120	& 6.545497	& 0.000930 \\
53867.835652	& 6.546418	& 0.001110 \\
53868.813512	& 6.547938	& 0.001090 \\
53869.789495	& 6.551369	& 0.001130 \\
53870.810097	& 6.549639	& 0.001290 \\
53871.815952	& 6.544530	& 0.001290 \\
53882.732970	& 6.543226	& 0.000960 \\
53886.703550	& 6.543739	& 0.000910 \\
53887.773514	& 6.543099	& 0.000860 \\
53917.737524	& 6.544316	& 0.001570 \\
53919.712544	& 6.548258	& 0.001610 \\
53921.615825	& 6.545739	& 0.000950 \\
53944.566259	& 6.545142	& 0.001460 \\
53947.578821	& 6.553284	& 0.002550 \\
53950.601834	& 6.548475	& 0.001390 \\
53976.497106	& 6.550300	& 0.001070 \\
53979.594316	& 6.546662	& 0.001620 \\
53981.555311	& 6.542953	& 0.001040 \\
53982.526504	& 6.543634	& 0.001170 \\
54167.866839	& 6.545060	& 0.001080 \\
54169.864835	& 6.547001	& 0.001110 \\
54171.876906	& 6.553622	& 0.001050 \\
54173.856452	& 6.546024	& 0.001090 \\
54194.847290	& 6.548596	& 0.001170 \\
54196.819157	& 6.544307	& 0.001270 \\
54197.797125	& 6.543237	& 0.001430 \\
54198.803823	& 6.543058	& 0.001270 \\
54199.854238	& 6.548529	& 0.000960 \\
54200.815699	& 6.549969	& 0.001040 \\
54201.918397	& 6.549210	& 0.001110 \\
54202.802697	& 6.543940	& 0.001140 \\
54227.831743	& 6.547675	& 0.001410 \\
54228.805860	& 6.551985	& 0.001120 \\
54229.773888	& 6.555056	& 0.001680 \\
54230.845843	& 6.549556	& 0.001100 \\
54231.801726	& 6.546737	& 0.001020 \\
54232.721251	& 6.545807	& 0.001780 \\
54233.910349	& 6.545118	& 0.001980 \\
54234.790981	& 6.545979	& 0.001140 \\
54253.728334	& 6.549170	& 0.001330 \\
54254.755898	& 6.544250	& 0.000980 \\
54255.709350	& 6.544921	& 0.001200 \\
54256.697674	& 6.547251	& 0.001320 \\
54257.704446	& 6.549922	& 0.001180 \\
54258.698322	& 6.552382	& 0.001120 \\
54291.675565	& 6.542981	& 0.001740 \\
54292.655662	& 6.545652	& 0.001250 \\
54293.708786	& 6.549893	& 0.001070 \\ \hline
\end{tabular}
\end{table}

\begin{table}
\caption{Radial-velocity measurements and errors bars for Gl~667C
  (first part). All 
  values are relative to the solar system barycenter, the secular
  acceleration is not substracted (see Sect.~2).}
\begin{tabular}{lll} \hline \hline
JD-2400000 & RV & Uncertainty \\ 
           & [km.s$^{-1}$] & [km.s$^{-1}$] \\ \hline
54295.628628	& 6.551364	& 0.001490 \\
54296.670395	& 6.543394	& 0.001230 \\
54297.631678	& 6.542745	& 0.001020 \\
54298.654206	& 6.541005	& 0.001200 \\
54299.678909	& 6.545336	& 0.001400 \\
54300.764649	& 6.547517	& 0.001120 \\
54314.691809	& 6.544615	& 0.002400 \\
54315.637551	& 6.549255	& 0.001840 \\
54316.554926	& 6.553586	& 0.001570 \\
54319.604048	& 6.540627	& 0.001070 \\
54320.616852	& 6.541618	& 0.001190 \\
54340.596942	& 6.546989	& 0.001030 \\
54342.531820	& 6.545501	& 0.001010 \\
54343.530662	& 6.549111	& 0.001110 \\
54346.551084	& 6.546453	& 0.001590 \\
54349.569500	& 6.543075	& 0.001270 \\
54522.886464	& 6.546384	& 0.001100 \\
54524.883089	& 6.550905	& 0.001190 \\
54525.892144	& 6.548526	& 0.001070 \\
54526.871196	& 6.548147	& 0.000990 \\
54527.897962	& 6.544307	& 0.001140 \\
54528.903672	& 6.542928	& 0.001240 \\
54529.869217	& 6.547428	& 0.001110 \\
54530.878876	& 6.549299	& 0.001000 \\
54550.901932	& 6.541460	& 0.001060 \\
54551.868783	& 6.544421	& 0.001000 \\
54552.880221	& 6.547682	& 0.000960 \\
54554.846366	& 6.549923	& 0.001180 \\
54555.870790	& 6.545093	& 0.001070 \\
54556.838936	& 6.544144	& 0.000980 \\
54557.804592	& 6.543654	& 0.001090 \\
54562.905075	& 6.548637	& 0.001050 \\
54563.898808	& 6.546148	& 0.001000 \\
54564.895759	& 6.544538	& 0.001200 \\
54568.891702	& 6.552451	& 0.001480 \\
54569.881078	& 6.547351	& 0.001270 \\
54570.870766	& 6.546062	& 0.001360 \\
54583.933324	& 6.547979	& 0.001560 \\
54587.919825	& 6.545972	& 0.001550 \\
54588.909632	& 6.550822	& 0.001510 \\
54590.901964	& 6.551713	& 0.001400 \\
54591.900611	& 6.547114	& 0.001430 \\
54592.897751	& 6.545115	& 0.001120 \\
54593.919961	& 6.544525	& 0.001140 \\
54610.878230	& 6.557495	& 0.001430 \\
54611.856581	& 6.554205	& 0.001010 \\
54616.841719	& 6.553078	& 0.001360 \\
54617.806576	& 6.554239	& 0.001740 \\
54618.664475	& 6.554319	& 0.002630 \\
54639.867730	& 6.551852	& 0.001550 \\
54640.723804	& 6.554482	& 0.001150 \\
54641.766933	& 6.550433	& 0.001270 \\
54642.676950	& 6.549163	& 0.001090 \\
54643.686130	& 6.543114	& 0.001250 \\
54644.732044	& 6.546094	& 0.001030 \\
54646.639658	& 6.552565	& 0.001430 \\
54647.630210	& 6.551306	& 0.001210 \\
54648.657090	& 6.548607	& 0.001420 \\
54658.650838	& 6.544032	& 0.001370 \\
54660.650214	& 6.547434	& 0.001520 \\
54661.760056	& 6.549334	& 0.001340 \\
54662.664144	& 6.550485	& 0.001440 \\
54663.784376	& 6.546365	& 0.001280 \\
54664.766558	& 6.547196	& 0.002130 \\
54665.774513	& 6.544636	& 0.001340 \\
54666.683607	& 6.547787	& 0.001290 \\
54674.576462	& 6.554292	& 0.001780 \\ \hline
\end{tabular}
\end{table}

\begin{table}
\caption{Radial-velocity measurements and errors bars for Gl~667C
  (third part). All 
  values are relative to the solar system barycenter, the secular
  acceleration is not substracted (see Sect.~2).}
\begin{tabular}{lll} \hline \hline
JD-2400000 & RV & Uncertainty \\ 
           & [km.s$^{-1}$] & [km.s$^{-1}$] \\ \hline
54677.663487	& 6.556773	& 0.003150 \\
54679.572671	& 6.547294	& 0.001850 \\
54681.573996	& 6.551406	& 0.001550 \\
54701.523392	& 6.548027	& 0.001160 \\
54708.564794	& 6.546551	& 0.001290 \\
54733.487290	& 6.557735	& 0.003860 \\
54735.499425	& 6.545217	& 0.001630 \\
54736.550865	& 6.542707	& 0.001520 \\
54746.485935	& 6.542513	& 0.000990 \\
54992.721062	& 6.555664	& 0.001140 \\
54995.741739	& 6.548206	& 0.001310 \\
54998.708975	& 6.552168	& 0.001460 \\
55053.694541	& 6.543819	& 0.001550 \\
55276.882590	& 6.549938	& 0.001260 \\
55278.827303	& 6.549239	& 0.001340 \\
55280.854800	& 6.551840	& 0.001360 \\
55283.868014	& 6.548172	& 0.001090 \\
55287.860052	& 6.553284	& 0.001120 \\
55294.882720	& 6.555268	& 0.001060 \\
55295.754277	& 6.556069	& 0.001450 \\
55297.805750	& 6.551650	& 0.001190 \\
55298.813775	& 6.550780	& 0.001150 \\
55299.785905	& 6.552531	& 0.002380 \\
55300.876852	& 6.553712	& 0.001240 \\
55301.896438	& 6.556232	& 0.001730 \\
55323.705436	& 6.557295	& 0.001270 \\
55326.717047	& 6.548716	& 0.001680 \\
55328.702599	& 6.550098	& 0.001470 \\
55335.651717	& 6.551042	& 0.001640 \\
55337.704618	& 6.554643	& 0.001680 \\
55338.649293	& 6.561893	& 0.002660 \\
55339.713716	& 6.553104	& 0.001920 \\
55341.789626	& 6.547775	& 0.001200 \\
55342.720036	& 6.554016	& 0.001580 \\
55349.682257	& 6.552810	& 0.001440 \\
55352.601155	& 6.562211	& 0.001750 \\
55354.642822	& 6.555963	& 0.001140 \\
55355.576777	& 6.553083	& 0.001430 \\
55358.754723	& 6.557765	& 0.001710 \\
55359.599377	& 6.554845	& 0.001450 \\
55673.791183	& 6.553326	& 0.001470 \\
55777.715412	& 6.562886	& 0.001990 \\
55779.530103	& 6.559617	& 0.001290 \\
55809.547632	& 6.553884	& 0.001180 \\
55815.538689	& 6.550158	& 0.001250 \\ \hline
\end{tabular}
\end{table}

\end{document}